\documentclass[twocolumn]{autart}
\usepackage{amsmath,amssymb,amsfonts,bm,xcolor}
\usepackage{algorithm,algorithmic,url}
\usepackage{multirow,rotating,booktabs}
\usepackage{savesym}
\usepackage{soul}
\usepackage{subcaption}
\savesymbol{AND}%

\newcommand{\algrule}[1][.2pt]{\par\vskip.5\baselineskip\hrule height #1\par\vskip.5\baselineskip}

\newcommand{\dint}{\int \!}
\newcommand{\dx}[1]{\,\mathrm{d}#1}

\definecolor{darkred}{RGB}{139,0,0}

\newcommand{\pmat}[1]{\begin{bmatrix}#1 \end{bmatrix}}
\newcommand{\Transp}{\mathsf{T}}
\newcommand{\ham}{\mathcal{H}}
\newcommand{\pot}{\mathcal{V}}
\newcommand{\kin}{\mathcal{K}}
\newcommand{\myVec}[1]{\textnormal{vec}\left ( #1 \right )}

\newtheorem{theorem}{Theorem}[section]
\newtheorem{lemma}{Lemma}[section]

\begin{document}
\begin{frontmatter}
	\runtitle{}
	\title{Practical Bayesian System Identification\\ using
          Hamiltonian Monte Carlo}

	\author[school]{Johannes Hendriks}
	\author[school]{Adrian Wills},
	\author[school]{Brett Ninness}
	\author[school]{Johan Dahlin},

	\address[school]{School of Engineering, The University of Newcastle, Callaghan NSW 2308, Australia.}


	\begin{abstract}
          This paper considers Bayesian parameter estimation of dynamic systems using a
          Markov Chain Monte Carlo (MCMC) approach.  The Metroplis--Hastings (MH)
          algorithm is employed, and the main contribution of the
          paper is to examine and illustrate the efficacy of a
          particular proposal density based on energy preserving Hamiltonian
          dynamics, which results in what is known in the statistics
          literature as ``Hamiltonian Monte--Carlo'' (HMC).  The very
          significant utility of this approach is that, as will be illustrated, it greatly
          reduces (almost to the point of elimination) the typically very high
          correlation in the Metropolis--Hastings chain which has been
          observed by several authors to restrict the application of
          the MH approach to only very low dimension model
          structures.  The paper illustrates how the HMC approach may
          be applied to both significant dimension linear and nonlinear model structures,
          even when the system order is unknown, and using both simulated
          and real data.
	\end{abstract}
\end{frontmatter}

\section{Introduction}
\label{sec:intro}
In many important applications, decisions must be made in the presence
of noise corrupted data and dynamic system model uncertainty. The
impact of these decisions ranges from trivial to catastrophic such as
the instability of a control loop.  Consequently, the ability to
manage uncertainty is vital, and this paper is directed at this
problem by illustrating how when dynamic system models are determined
from noise corrupted data, the associated estimated system uncertainty
can be quantified.

Furthermore, we illustrate how the uncertainty in {\em functions} of the
estimated dynamic system, such as control system performance
measures, e.g. stability margin for a putative control law, can also be quantified.

The topic of estimating dynamic system models from observed data has been
studied in earnest for half a century, with the overwhelming majority
of solutions being highly effective parametric methods based on optimisation techniques
such as least squares (LS) and maximum likelihood (ML) and
generalisation such as the prediction error (PE) approach.

These approaches have their origins in the statistics literature.
Except for very simple cases, such as linearly parameterised model
structures, quantifying the error in the associated parameter
estimates that arise from stochastic data corruption relies on the employment
of a central limit theorem (CLT).  Unfortunately, this gives an error
quantification that is approximate, since it relies on assuming the
asymptotic in data length $N\rightarrow\infty$ CLT quantification is
approximately true for finite data length $N$.  Furthermore, it is
often very challenging to apply this CLT-based approach to nonlinear
dynamic system estimation.

This paper is motivated by these difficulties, particularly that of
accurate error quantification for nonlinear systems and/or small data
length $N$, and examines the utility of a solution based on a Bayesian
methodology.  The rationale for this approach is that error bounds
based on Bayesian posterior densities are non-asymptotic in data
length $N$, and the ability to inject quite sophisticated prior
knowledge can significantly improve the accuracy of the obtained
estimates.

Despite these attractive features, the Bayesian approach is not
without its challenges.  Foremost is that it is an approach that is
not based on numerical optimisation methods that underpin LS, ML and
PE methods, but instead requires
numerical {\em integration} methods.  In very special cases such as
the linearly parametrised linear Gaussian model with so-called
``conjugate'' priors, the associated integration can be provided in
closed form, and in more general cases of very low model dimension the
required integrals can be computed numerically using, for example,
quadrature methods.

However, in this latter case the number of function evaluations
required grows as a parameter-dimension {\em exponential} of the
number of integration ``bins'', and hence grows dramatically in both.
Furthermore, the range of integration for each parameter to be
estimated must be specified, which for the problems considered here is
impossible since the parameters themselves are unknown a priori.  This
renders this numerical integration approach inapplicable except in
very special very low parameter dimension situations.

An alternative numerical integration approach studied in this paper is
based on constructing a random number generator for which the samples
are distributed according to the Bayesian posterior distribution. The
utility of this random number generator is that sample-averages of the
random number generator realisations can be used to approximate, with
arbitrary accuracy for sufficiently large number of samples, the
required integrals.  For example, it can provide estimates of the
conditional mean, which is known to be the minimum mean-squared
estimate.

Constructing such a random number generator has received significant
research attention and among the possible solutions is the so-called
Markov-Chain approach. As the name suggests, this approach constructs
a Markov-Chain whose stationary distribution coincides with desired
posterior density target distribution.  Ensuring that this
Markov-Chain does indeed converge to the desired stationary
distribution can be achieved using the decades old and ingenious
Metropolis--Hastings (MH) algorithm which dates back to the 1950's
were it was developed to compute problems in physical chemistry~\cite{Metropolis:1949,Metropolis:1953}.

Despite achieving this remarkable outcome, the MH method has a
significant drawback in that it requires a certain user-supplied
``proposal density''. This proposal must be carefully selected to
deliver Markov chain realisations that are sufficiently uncorrelated
that the approach delivers sufficiently accurate computation of
desired integrals without requiring an unreasonably large amount of
realisations being generated.  Historically, this practical aspect has
prohibited the application of the MH algorithm to dynamic system
estimation of all but the simplest model structures.

This paper addresses this weakness by employing the so-called ``Hamiltonian
 Monte Carlo'' approach, which uses a particular proposal density
 that, as will be illustrated here, renders the MH-algorithm far less
 sensitive to user tuning and scales far more successfully to higher dimensions. 

 Hamiltonian Monte Carlo (HMC) is a relatively recent development
 originating in the physics community~\cite{Duane} and then taken up
 by the statistics community~\cite{GirolamiCalderhead2011,Neal2010}
 and now is attracting interest in machine learning~\cite{Betancourt2017}.

However, despite its great potential, it appears to be largely
unrepresented in the control community.  This paper is directed at
filling this gap by introducing the HMC approach in applications ranging
from classical linear transfer function models through to quite
general nonlinear state-space models.  Furthermore estimation of models of a-priori unknown
model order and estimation together with error quantification of {\em
  functions} of the parameter estimates are also illustrated.


\section{Problem Formulation} 
\label{sec:problem-description}

The problem addressed here is that of using measured data from a dynamic system consisting of inputs
$u_t$ and outputs $y_t$ for $t=1,\ldots,N$ to determine a
model for the system.

Typically, such a model will have several
unknown parameters which we collect in a vector
$\theta\in \mathbf{R}^{n_{\theta}}$ and will typically be a function of
past inputs and outputs plus also potentially latent variables, such
as system states $x_t$.

A very general mathematical description of this model could be written as
\begin{equation}
	y_{t+1} = f(\theta,y_{1:t},u_{1:t+1},x_{1:t},e_{1:t+1}),
\end{equation}
where we use the shorthand $y_{1:t}$ to denote the set
$\{y_1,\ldots,y_t\}$ etc.\ , $x_t$ is a possible state vector, and $e_t$ is an unobserved stochastic
sequence where all realisations are independent and
identically distributed (i.i.d).

In this paper, the function $f$ can be quite
general, but in simple cases could take the form of, for example, a
transfer function or linear state-space model.

The aim of this work is to evaluate expectation integrals with respect
to a posterior distribution $p(\theta\mid y_{1:t})$ of the form 
\begin{equation}
  \label{eq:expectation_int}
	I = \mathbf{E}\{\varphi(\theta) | y_{1:N}\} = \int \varphi(\theta) p(\theta | y_{1:N})\,\mathrm{d}\theta,
\end{equation}
where $\varphi:\theta \to \mathbf{C}^n$ is a general function of the
parameters, such as the conditional mean, Bode frequency response, or
other almost arbitrary functions such as achieved phase margin for a
given controller.



Throughout this paper, while the vector
$\theta\in\mathbf{R}^{n_{\theta}}$ denotes the model parameters the
vector $\eta\in\mathbf{R}^{n_{\eta}}$ with $n_{\eta}\geq n_{\theta}$
will be used to denote the full vector of elements we wish to draw samples
from using MCMC. It is important to make this distinction as for some
problems $\eta$ could also include values additional to $\theta$ which
are latent variables, such as the system state history  $x_{1:N}$ or
quantification of any noise sequences such as their variances.


\section{A Markov Chain Monte Carlo Approach} 
\label{sec:markov-chain-monte}

The computational approach studied here relies on the Law of Large Numbers
to provide estimates of \eqref{eq:expectation_int} via so-called
Monte-Carlo integration where samples
\begin{equation}
  \label{eq:5}
\eta_{k}\sim p(\eta \mid y_{1:N}),\qquad i=k\cdots M 
\end{equation}
are used to compute a sample average of the integral $I$ in
\eqref{eq:expectation_int} according to
\begin{align}
  \label{eq:3}
I\approx  \widehat{I}_M = \frac{1}{M} \sum_{k=1}^M \varphi(\eta_{k}).
\end{align}
Under rather general assumptions on $p(\eta \mid y_{1:N})$ and
$\varphi(\cdot)$, it can be shown that 
\begin{align}
  \label{eq:4}
  \widehat{I}_M \rightarrow I \quad \text{as} \quad  M\to \infty
\end{align}
where the convergence is with probability $1$.  The ``Monte Carlo''
epithet is in reference to the famous casino, and a nod to the random
aspects of the approach.

Note that $\varphi(\cdot)$ may be chosen as the indicator function which is
equivalent to $\widehat{I}_M$ being an estimate of the posterior
density $p(\eta\mid y_{1:t})$, or marginals of it,
via a sample histogram of the realisations of all of $\{\eta_{i}\}$ or
the component of it for which a marginal is of interest.

Importantly, the rate of convergence in \eqref{eq:4} depends only on
the correlation between the realisations $\{\varphi(\eta_i)\}$ and {\em not}
on the dimension $n_{\eta}$ of $\eta$, thus removing the exponential
computational load problem associated with quadrature and other deterministic
methods mentioned in the introduction.

Furthermore, bounds for integration limits need not be
specified a-priori since the realisations naturally arise mainly in
regions that significantly affect the evaluation of $I$ and rarely in
regions that do not.

However, the ability to use this approach is predicated on being able
to draw the samples~\eqref{eq:5}.  A remarkably simple and elegant approach to
solve this problem is to employ the so-called ``Metropolis--Hastings''
(MH) algorithm.



\subsection{The Metropolis-Hastings Algorithm}
\label{sub:MH}

The Metropolis--Hastings algorithm simulates a Markov chain using two steps at each iteration
$k$.  First, a candidate parameter $\eta'$ is simulated using a
\textit{proposal distribution} $q$ by
\begin{align}
	\eta'\sim q(\eta' | \eta_{k-1}),\label{eq:mh:proposal}
\end{align}
where $q$ is selected by the user.
A standard choice is a Gaussian proposal given by
\begin{align}\label{eq:mh:gaussian:proposal}
	q(\eta' | \eta_{k-1}) = \mathcal{N} \left(\eta';\mu(\eta_{k-1}),\Sigma(\eta_{k-1})\right),
\end{align}
A common very simple specification in this case is $\mu(\eta_{k-1}) = \eta_{k-1}$ and $\Sigma(\eta_{k-1}) =
\epsilon^2 I_{n_{\eta}}$ for some user-tuned value of $\epsilon$.

Second, a desired ``target'' density $\pi(\eta)$ from which it is
desired to draw realisations is specified, such as
$\pi(\eta)=p(\eta\mid y_{1:N})$ and then
the candidate parameter is accepted with probability given by
\begin{align} \label{eq:mh:aprob}
	\alpha(\eta^{\prime}, \eta_{k-1}) =\min\left\{1,\frac{\pi(\eta^{\prime})}{\pi(\eta_{k-1})}\frac{q(\eta_{k-1} | \eta')}{q(\eta^{\prime} | \eta_{k-1})}\right\}.
\end{align}
The final step is to set $\eta_k \leftarrow \eta'$ when the candidate is accepted and
$\eta_k \leftarrow \eta_{k-1}$ when it is rejected.  The full
procedure is presented in Algorithm~\ref{alg:mh}.

In essence this approach is such that an arbitrary Markov
chain that the user can easily simulate, as specified by the
``proposal distribution'' $q(\eta_{k}\mid\eta_{k-1})$ is then
modified to a Markov chain whose stationary density is the required
target $\pi(\eta)$.

Note that is common that the proposal is symmetric (eg,
$\eta^{\prime}\sim\mathcal{N}(\eta,\epsilon I))$ in which case
$q(\eta_{k-1} | \eta')/q(\eta' | \eta_{k-1})=1$ and the acceptance
probability simplifies to
\begin{align} \label{eq:m:aprob}
	\alpha(\eta^{\prime}, \eta_{k-1}) =\min\left\{1,\frac{  \pi(\eta^{\prime}) }{ \pi(\eta_{k-1}) }\right\}.
\end{align}
in which case the resultant algorithm is known simply as the
``Metropolis'' method.

\begin{algorithm}[!t]
	\caption{\textsf{Metropolis-Hastings (MH)}}
	\footnotesize
	\textsc{Inputs:} $M>0$, $\eta_0$ and $q(\cdot\mid\cdot)$.\\ \textsc{Output:} $\{\eta_k\}_{k=1}^K$.
	\algrule[.4pt]
	\begin{algorithmic}[1]
		\STATE Compute $\pi(\eta_0)$.
		\FOR{$k=1$ to $M$}
		\STATE Sample $\eta' \sim q(\eta' | \eta_{k-1})$ using \eqref{eq:mh:proposal}.
		\STATE Sample $\omega_k$ uniformly over $[0,1]$.
		\IF{$\omega_k \leq \min\{1,\alpha(\eta',\eta_{k-1})\}$ given by \eqref{eq:mh:aprob}.}
		\STATE Accept $\eta'$, i.e.\
		$\eta_k \leftarrow \eta'$.
		\ELSE
		\STATE
		Reject $\eta'$, i.e.\
		$\eta_k \leftarrow \eta_{k-1}$.
		\ENDIF
		\ENDFOR
	\end{algorithmic}
	\label{alg:mh}
  \end{algorithm}



\subsection{Sufficient Invariance Condition - Detailed Balance}
\label{sec:suff-cond-deta}

It is important to understand the fundamentals of the design of the
MH algorithm.   First, observe that it generates a time-homogeneous
Markov chain with transition kernel $K(\eta_k\mid \eta_{k-1})$ given
by
\begin{equation}
  \label{eq:1}
K(\eta_k=\xi_{k}\mid \eta_{k-1}) =  \alpha(\xi_k\mid\eta_{k-1} )q(\xi_{k}\mid\eta_{k-1})
\end{equation}
where, to avoid unnecessary technicalities that may obscure intuition,
we neglect the very rare event that drawing from the proposal
$q(\cdot\mid\eta_{k-1})$ for $\eta_{k}$ would deliver {\em exactly} the same value
$\eta_{k-1}$ as in the previous iteration. 

Therefore, the probability density $\pi_k$ for the $k$'th iteration of
the chain, is related to the one $\pi_{k-1}$ at the previous iteration
according to the law of total probability as
\begin{equation}
  \label{eq:2}
  \pi_k(\eta_{k}) = \int K(\eta_k\mid
  \eta_{k-1})\,\pi_{k-1}(\eta_{k-1})\,{\rm d}\eta_{k-1}
\end{equation}
and therefore, if the chain is to converge to deliver realisations
from a stationary distribution $\pi$ then the latter must satisfy the
equation
\begin{equation}
  \label{eq:6}
 \pi(\eta) = \int K(\eta\mid \xi)\,\pi(\xi)\,{\rm d}\xi 
\end{equation}
in which case $\pi$ is referred to as an ``invariant'' distribution of
the chain.  

The MH algorithm is specifically designed so that the kernel
$K(\cdot\mid\cdot)$ it implements has the target $\pi$ as an invariant
distribution by ensuring the chain it simulates satisfies a condition
which is sufficient (but not necessary) for invariance, which is
termed the ``detailed balance'' condition:
\begin{equation}
K(\eta\mid\xi)\pi(\xi) = 
K(\xi\mid\eta)\pi(\eta).
\label{eq:26}
\end{equation}
To see the sufficiency of this condition for invariance, note that it implies
\begin{eqnarray}
\int K(\eta\mid\xi) \pi(\xi)\,{\rm d}\xi &=& 
\int K(\xi\mid \eta) \pi(\eta )\,{\rm d}\xi\nonumber\\
&=& \pi(\eta )\int K(\xi\mid \eta) \,{\rm
  d}\xi= \pi(\eta ).
\qquad \label{eq:16a}
\end{eqnarray}
To see that the very simple design of the MH algorithm ensures
detailed balance, we compute
\begin{align}
\pi(\eta)K(\xi\mid\eta) 
&=\pi(\eta)
q(\xi\mid\eta) 
\min\left\{1,
\frac{\pi(\xi)}{\pi(\eta)}
\cdot
\frac{q(\eta\mid\xi)}{q(\xi\mid\eta)}
\right\}\nonumber \\
&= \min\left\{
\pi(\eta)
q(\xi\mid\eta),
\pi(\xi)
q(\eta\mid\xi)
\right\}.\label{eq:24}
\end{align}
Similarly, 
\begin{align}
\pi(\xi)K(\eta\mid\xi) 
&=
\pi(\xi\mid ,Y)
q(\eta\mid\xi) 
\min\left\{1,
\frac
{\pi(\eta)}
{\pi(\xi)}
\cdot
\frac
{q(\xi\mid\eta)}
{q(\eta\mid\xi)}
\right\}\nonumber \\
&= \min\left\{
\pi(\xi)
q(\eta\mid\xi)
,
\pi(\eta)
q(\xi\mid\eta)
\right\}\label{eq:25}.
\end{align}
Comparing (\ref{eq:24}) and (\ref{eq:25}) and noting that the
$\min\{\cdot,\cdot\}$ operation is symmetric establishes that the
 ``detailed balance'' condition\eqref{eq:26} and hence the invariance
 condition \eqref{eq:6} is ensured.

For more extensive detail on this and related MH convergence analysis
the interested reader is referred to
\cite{Tierney1994,NinnessHenriksen2010,RobertCasella2004}.

\subsection{Constructing proposal distributions}
\label{sec:practice}

The proposal distribution $q(\cdot\mid \cdot) $ is the only tuning
aspect of the MH algorithm and determines its performance
in terms of how quickly the accuracy of the approximation
\eqref{eq:3} improves with increasing number of realisations $L$.

One way of judging this rate is to employ the Central Limit Theorem
(CLT) for the estimate \eqref{eq:3} which is given by~\cite{RobertCasella2004}:
\begin{equation}
  \label{eq:8}
  \sqrt{M}\left[ \widehat{I}_M - I
\right]
\stackrel{\mathcal{D}}{\longrightarrow}\mathcal{N}(0,\sigma^2_{\varphi}),\qquad L\rightarrow\infty
\end{equation}
where
\begin{equation} 
	\sigma^2_{\varphi} \propto\mathsf{IACT}(\{\varphi(\eta_k)\})
    =1+2\sum_{l=1}^{\infty}\mathsf{corr}(\varphi(\eta_0), \varphi(\eta_k))
\label{eq:mh:asymptotic:variance}
\end{equation}
where the integrated auto-correlation time (IACT) is introduced and
infers the convergence rate approximation
\begin{equation}
\label{eq:Lapprox}
|\widehat{I}_M-I|^2\propto \frac{\sigma^2_\varphi}{M}.
\end{equation}
On the one hand, this result emphasises the great attraction of this
MCMC method for computing integrals since the $1/L$ convergence rate
in~\eqref{eq:Lapprox} is \emph{independent} of
the dimension of $\eta$ which determines the dimension of any
integrals involved.

On the other hand, the dimension of $\eta$ is still an issue since it
is typically observed that as it grows, the complexity of the target
$\pi(\eta)$ grows, and it becomes increasingly difficult to design a
proposal $q(\cdot\mid\cdot)$ such that the correlation $\mathsf{corr}(\varphi(\eta_0), \varphi(\eta_k))$
decreases quickly with increasing $k$.

For example, for a proposal $q(\cdot\mid\cdot)$ with large steps designed to
explore $\pi(\cdot)$ well, one or more elements of $\eta$ are often moved
well away from the support of $\pi(\cdot)$ and draw
$\alpha(\cdot\mid\cdot)$ very low so that almost all steps are rejected, and
hence the chain stays in the same place for very long periods and is
highly correlated.

Vice versa, if only tiny steps are proposed by $q(\cdot\mid\cdot)$ in
order to ensure many acceptances, the chain wanders very slowly, and
again in a highly correlated manner akin to a random walk.

Both of these generate a very high, perhaps infinite value for
$\sigma^2_{\varphi}$ in \eqref{eq:Lapprox}.  The literature on MCMC
therefore concentrates strongly on the so-called ``mixing'' of the
chain with ``strong mixing'' meaning a quick decrease in
$\mathsf{corr}(\varphi(\eta_0), \varphi(\eta_k))$ with increasing $k$.

One commonly used approach to address this issue is to use the
proposal choice~\eqref{eq:mh:gaussian:proposal} with the choices
\begin{equation}
  \label{eq:mh:gaussian:proposal:ruleofthumb}%
	\mu(\eta_{k-1}) =\eta_{k-1}, \qquad
	\Sigma(\eta_{k-1}) = \widehat{\Sigma}_{\pi},
 \end{equation}
where $\widehat{\Sigma}_{\text{post}}$ denotes an estimate of the
target distribution covariance.  The rationale here is that the support of a
Gaussian with these choice for mean and covariance approximates the
support of the target $\pi$.

Typically, this is obtained by starting with a covariance
$\Sigma(\eta_{k-1})=\epsilon^{2}I$ for some choice of $\epsilon$, and
then after an initial pilot run $\{\eta\}_k$ has been obtained,
computing $\Sigma_{\pi}$ via a sample average estimate from this run.

Often, this involves trial and error choice of $\epsilon$ to
achieve sufficient mixing of the pilot run, which in many cases of
non-trivial dimension is not possible.

Another approach for constructing a proposal is to borrow ideas from
the non-linear optimisation literature and employ a quasi-Newton
optimisation step as a proposal by again using the Gaussian proposal
choice~\eqref{eq:mh:gaussian:proposal} with
\begin{subequations}\label{eq:mh:gaussian:proposal:newton}%
\begin{align}
	\mu(\eta_{k-1})& =\eta_{k-1}+\frac{\epsilon^2}{2} H^{-1}(\eta_{k-1}) G(\eta_{k-1}),    \\
	\Sigma(\eta_{k-1})& = \epsilon^2 H^{-1}(\eta_{k-1}).
\end{align}%
\end{subequations}%
Here, we introduce the gradient of the log-posterior,
\begin{align*}
	G(\eta')= \nabla \log \pi(\eta) \big|_{\eta=\eta'},
\end{align*}
and the negative inverse Hessian of the log-posterior
\begin{align*}
	H^{-1}(\eta') =- \nabla^2 \log \pi(\eta) \big|_{\eta=\eta'}.
\end{align*}
This strategy takes into account the gradient and curvature of the
target distribution and therefore introduces
a mode-seeking behaviour, which guides the Markov chain to areas of
higher likelihood.

We refer to this type of proposal as the manifold Metropolis adjusted
Langevin algorithm (mMALA) as it is known in the statistics
literature, see \cite{GirolamiCalderhead2011} where it has been
successfully applied to, for example, econometric problem with 
up to 20 states and 50+ parameters)~\cite{AdolfsonLasenLindeVillani2007b}.

While these proposal choices
\eqref{eq:mh:gaussian:proposal:ruleofthumb} and
\eqref{eq:mh:gaussian:proposal:newton} are well accepted as improving
the performance of the MH method, they fall short of achieving the
optimum of delivering a sequence of realisations $\{\eta_k\}$ that are
 both uncorrelated, and therefore also involve few rejections.  The
Hamiltonian Monte Carlo Method (HMC) is designed to achieve these
latter goals.

\section{Hamiltonian Monte Carlo} 
\label{sec:basics_of_hamiltonian_monte_carlo}

The essential idea of Hamiltonian Monte Carlo (HMC) is to augment the
parameter space by adding an additional vector $\rho$ where 
\begin{equation}
\label{eq:11}
\mbox{dim} (\rho) = \mbox{dim} (\eta) = n_{\eta}
\end{equation}
and then employ Metropolis Hastings to generate samples from a new
augmented target density 
\begin{equation}
\label{eq:16}
\widetilde{\pi}(\eta,\rho) = \pi(\eta\mid\rho)p(\rho)
\end{equation}
for some user chosen density $p(\rho)$.
In this case, the marginal of $\widetilde{\pi}(\eta,\rho)$ with respect to $\rho$ is the target $\pi(\eta)$ of
interest by the law of total probability:
\begin{equation}
\label{eq:12}
\int \widetilde{\pi}(\eta,\rho)\,{\rm d}\rho= \int \pi(\eta\mid\rho)
p(\rho)\,{\rm d}\rho = \pi(\eta).
\end{equation}

In HMC, the ``augmented'' density $\widetilde\pi$ is formulated as
\begin{equation}
\label{eq:13}
	\widetilde{\pi}(\eta,\rho) = \exp(-H(\eta,\rho)), 
\end{equation}
where 
\begin{equation}
  \label{eq:15}
  \ham(\eta,\rho) = -\log\widetilde{\pi}(\eta,\rho) = \pot(\eta) + \kin(\rho)
\end{equation}
with
\begin{equation}
  \label{eq:18}
  \pot(\eta) =-\log \pi(\eta), \quad
  \kin(\rho) =-\log p(\rho\mid\eta).
\end{equation}
and $\ham(\eta,\rho)$ is interpreted as the Hamiltonian of a dynamic
system in $\eta(t),\rho(t)$ that evolves according to Hamilton's
equations:
\begin{equation}
\label{eq:17}
\frac{{\rm d}\eta_i(t)}{{\rm d}t} =\frac{\partial \ham(\eta,\rho)}{\partial \rho_i
},\qquad
\frac{{\rm d}\rho_i(t)}{{\rm d}t} =-\frac{\partial \ham(\eta,\rho)}{\partial \eta_i }
\end{equation}
so that $\eta$ can be imagined as a position vector and $\rho$ a
momentum vector.  In this case $\pot(\eta)$ can be considered the
potential energy and $\kin(\rho)$ the kinetic energy of a particle in
motion governed by \eqref{eq:17}.

Importantly, Hamiltonian dynamics are energy preserving as can be
simply established via
\begin{eqnarray}
\frac{{\rm d} \ham(\eta(t),\rho(t))}{{\rm d}t} &=&
\sum_{i=1}^{n_{\rho}}
\left[
\frac{{\rm}\eta_{i}(t)}{{\rm d}t}
\frac{\partial \ham}{\partial\eta_{i}}
 +
\frac{{\rm}\rho_{i}(t)}{{\rm d}t}
\frac{\partial \ham}{\partial\rho_{i}}
\right]\\
&=&
\sum_{i=1}^{n_{\rho}}
\left[
\frac{\partial \ham}{\partial \rho_i}
\frac{\partial \ham}{\partial\eta_{i}}
 -
\frac{\partial \ham}{\partial\eta_i}
\frac{\partial \ham}{\partial\rho_{i}}
\right] = 0.
\end{eqnarray}
Therefore, if we construct a MH algorithm for which a new realisation
$(\eta^{\prime},\rho^{\prime})$ is governed by the dynamics
\eqref{eq:17} then, if in addition the proposal $\gamma(\cdot\
\mid \cdot)$ is symmetric the acceptance probability for this new
realisation is
\begin{eqnarray}
\label{eq:20}
\alpha(\eta^{\prime},\rho^{\prime},\eta,\rho) &=& \min\left\{
1,\frac{\widetilde{\pi}(\eta^{\prime},\rho^{\prime})}{\widetilde{\pi}(\eta,\rho)}
\right\} \label{eq:hmc_acceptance}\\
&=&
\min\left\{
1,\exp\left( -\ham(\eta^{\prime},\rho^{\prime}) + \ham(\eta,\rho) \right)
\right\}
\nonumber \\
&=& \min\{1,\exp(0)\} = 1
\label{eq:22}
\end{eqnarray}
thus realising the ambition of a high acceptance rate.  Realising the
ambition of uncorrelatedness between $(\eta^\prime,\rho^\prime)$ and
$(\eta,\rho)$ can be achieved by drawing a random proposal $\rho \sim
p(\rho)$ independant of any previous iterations and then simulating the
Hamiltonian system over some user chosen time period $t\in[0,L]$ to deliver
$(\eta^{\prime},\rho^{\prime})$ independent of $(\eta,\rho)$.

This is the HMC algorithm, which for the choice of independently drawn
$\rho\sim p(\rho\mid\eta)=\mathcal{N}(0,\mathcal{M})$ which implies kinetic energy of
\begin{equation}
	\kin(\rho) = \frac{1}{2}\rho^{\Transp}\mathcal{M}^{-1}\rho +
    \log|\mathcal{M}| + \text{const},
    \label{eq:9}
\end{equation}
is formally defined in Algorithm~\ref{alg:hmc} with $\mathcal{M}$
referred to as a ``mass matrix'':
\begin{algorithm}[!t]
	\caption{\textsf{Hamiltonian Monte Carlo (HMC)}}
	\footnotesize
	\textsc{Inputs:} $M>0$, $\eta_0$, $\mathcal{M}$, $\epsilon$, $L$
    and $\pot=-\log\pi(\eta)$.\\ 
\textsc{Output:} $\{\eta_k\}_{k=1}^K$.
	\algrule[.4pt]
	\begin{algorithmic}[1]
		\FOR{$k=1$ to $M$}
		\STATE Sample $\rho_{k-1} \sim \mathcal{N}(0, \mathcal{M})$.
		\STATE $(\eta_L,\rho_L) \leftarrow \texttt{Leapfrog\_Integrator}(\eta_{k-1},\rho_{k-1},\epsilon,L,\pot,\mathcal{M})$, Algorithm .
		\STATE Set $(\eta',\rho') \leftarrow (\eta_L,-\rho_L)$ as the proposed sample.
		\STATE Sample $u$ uniformly over $[0,1]$.
		\IF{$u \leq \min\{1,\alpha(\eta',\rho',\eta_{k-1},\rho_{k-1})\}$ given by \eqref{eq:hmc_acceptance}.}
		\STATE Accept $(\eta',\rho')$, i.e.\
		$(\eta_k,\rho_k) \leftarrow (\eta',\rho')$.
		\ELSE
		\STATE
		Reject $(\eta',\rho')$, i.e.\
		$(\eta_k,\rho_k) \leftarrow (\eta_{k-1},\rho_{k-1})$.
		\ENDIF
		\ENDFOR
	\end{algorithmic}
	\label{alg:hmc}
\end{algorithm}

There are some important subtleties that require attention.  First,
the reader may note that the simulation step is a
deterministic transformation 
$\Phi(\eta,\rho)\mapsto (\eta^{\prime},\rho^{\prime})$ and hence a formal change of
variables calculation should be applied when computing the acceptance probability
\eqref{eq:20}.  This can be accomplished using the well know change of
variables formula~\cite{PAPOULIS} which would involve a multiplicative factor in
\eqref{eq:20} of $\det(\nabla_{\eta',\rho'})\Phi^{-1}(\eta^{\prime},\rho^{\prime}) $.

However, an important property of Hamiltonian dynamics is that they
are \emph{volume preserving}, so that this determinant is always equal
to one and can be neglected~\cite{Neal2010}.  

Furthermore, Hamilton's equations~\eqref{eq:17} are time reversible
in the sense that if they govern the movement of a particle for $L$ seconds,
if the momentum of that particle is reversed in sign, then after
another $L$ seconds it will return to where it originally was.  This
explains the negative sign on $\rho_L$ in step 4 of
Algorithm~\ref{alg:hmc} since it ensures the detailed balance
condition~\eqref{eq:26} holds and that $\widetilde\pi$ is an invariant
density for the resulting chain~\cite{Duane}.

Finally, the reader may also note that between lines 6 and 10 of
Algorithm~\ref{alg:hmc} the acceptance ratio is computed and used to
accept or reject, even though as established in~\eqref{eq:22} this
acceptance ratio should always be one, and hence no rejections.

The former accounts for the fact that numerical simulation of the
Hamiltonian system will necessarily involve some errors, hence exact
equality of the acceptance probability to unity cannot be expected.

Simulating the Hamiltonian system from $t=0$ to $t=L$ involves
computing the integrals
\begin{equation}
\label{eq:7}
\eta(L) = -\int_{0}^L \frac{\partial \kin(\rho(t))}{\partial\rho}\,{\rm
  d}t,\quad
\rho(L) = \int_{0}^L \frac{\partial \pot(\eta(t))}{\partial \eta}\,{\rm d}t
\end{equation}
Fortunately, there exists a class of so-called ``symplectic'' integrators
designed specifically for the simulation of Hamiltonian systems and to
be as close to volume preserving as possible.   One deceptively simple
one used in HMC applications is the ``leap frog'' method specified in
Algorithm~\ref{alg:leapfrog} again for the kinetic energy choice~\eqref{eq:9}
which is used at line 3 of the HMC Algorithm~\ref{alg:hmc}.
\begin{algorithm}[!t]
	\caption{\textsf{Leapfrog Integrator}}
	\footnotesize
	\textsc{Inputs:} $\eta$, $\rho$, $\mathcal{M}$, $\epsilon$, $L$ and $\pot$.\\ \textsc{Output:} $(\eta_L,\rho_L)$.
	\algrule[.4pt]
	\begin{algorithmic}[1]
		\STATE $\eta_0,\rho_0 \leftarrow (\eta,\rho)$
		\FOR{$0 \leq n < \text{floor}(L/\epsilon)$}
		\STATE $\rho_{n+1/2} \leftarrow \rho_n - \frac{\epsilon}{2}\frac{\partial \pot(\eta)}{\partial \eta}\Big|_{\eta=\eta_n}$.
		\STATE $\eta_{n+1} \leftarrow \eta_n + \epsilon \mathcal{M}^{-1}\rho$.
		\STATE $\rho_{n+1} \leftarrow \rho_{n+1/2} - \frac{\epsilon}{2}\frac{\partial \pot(\eta)}{\partial \eta}\Big|_{\eta=\eta_{n+1}}$
		\ENDFOR
		\STATE $(\eta_L,\rho_L) \leftarrow (\eta_{n+1},\rho_{n+1})$
	\end{algorithmic}
	\label{alg:leapfrog}
\end{algorithm}

\subsection{HMC Tuning Choices}
\label{sec:hmc-tuning-choices}

Unfortunately, there are some user defined choices that need to be
made to implement HMC.  Namely, the Hamiltonian simulation duration
$L$, the symplectic integration step size $\epsilon$ and the kinetic
energy mass matrix $\mathcal{M}$ which is the covariance of the Normal
distribution for new momentum proposals.

While $L$ and $\epsilon$ can be statically
defined~\cite{Betancourt2017}, a state-of-the-art implementations uses
a so called ``No-U-Turn'' criteria
\cite{HoffmanGelman2014,carpenter2017stan} to dynamically adjust the
integration time, and the integration step size is dynamically tuned
to achieve the desired acceptance rate.

Additionally, there is the choice of mass matrix $\mathcal{M}$ which
can be chosen by computing an empirical estimate of the target covariance
using a warm-up phase of the Markov chain.  This choice of mass matrix
decorrelates the parameters and gives more uniform energy levels.

However, unless the target distribution is exactly Gaussian no choice
of global rotation and scaling will yield completely uniform level
sets.


\subsection{Example}
\label{sec:example}

To illustrate the performance advantages of the HMC proposal in a MH
algorithm, we consider an example in a two-dimensional parameter space
$\eta = \{\eta_1,\eta_2\}$ where the target distribution $\pi(\eta)$
has a doughnut-shaped support.

This may seem contrived, however, it is recognised that many common
targets in higher dimensional spaces tend to occupy a
narrow shell~\cite{Betancourt2017}.

\begin{figure*}[htb]
\centering
\subcaptionbox{MH $k=1$\label{fig:donut_MH_1}}
{\includegraphics[trim={10 5 45 25},clip,width=0.4\linewidth]{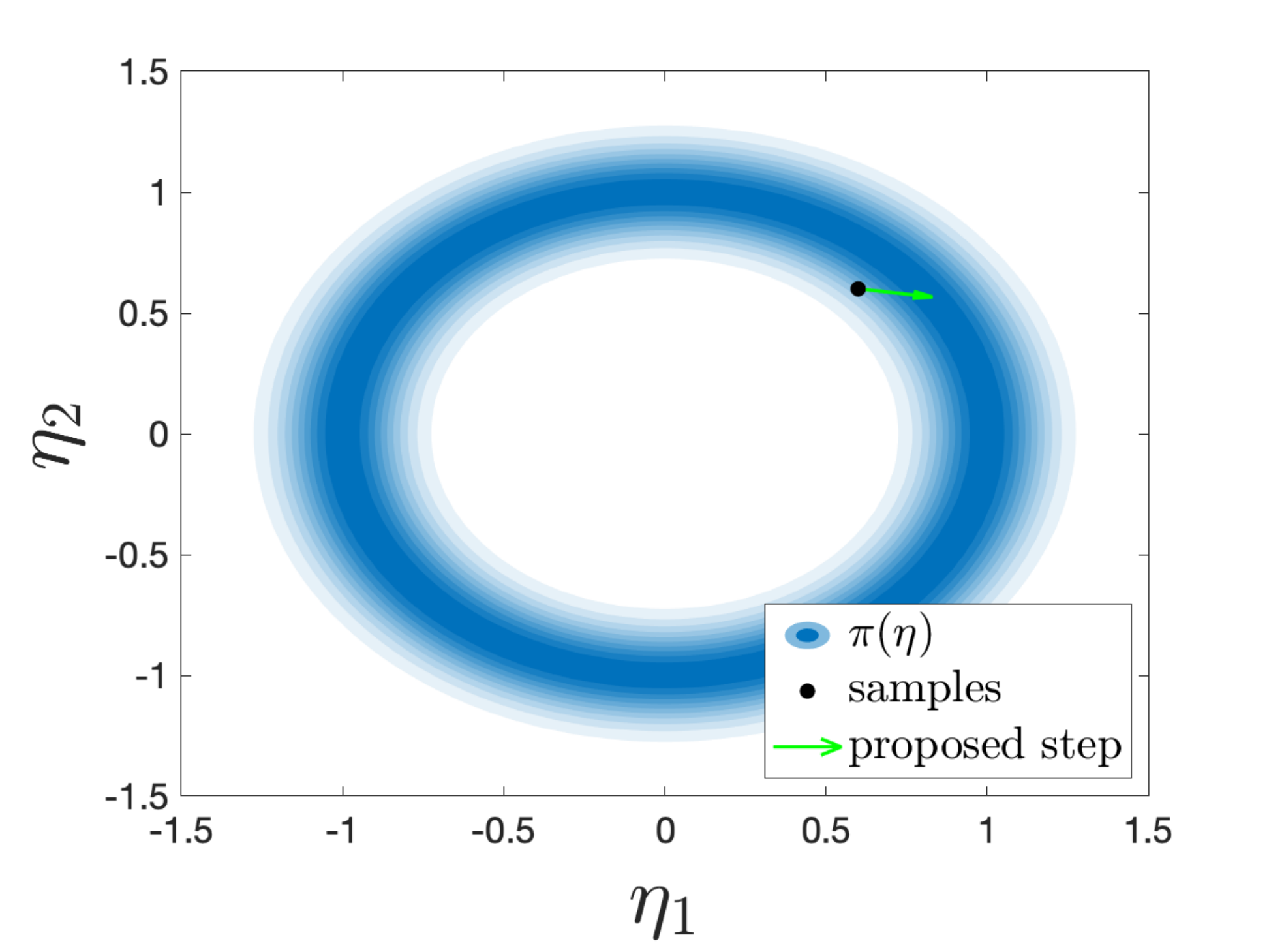}}
\subcaptionbox{MH $k=2$\label{fig:donut_MH_2}}
{\includegraphics[trim={10 5 45 25},clip,width=0.4\linewidth]{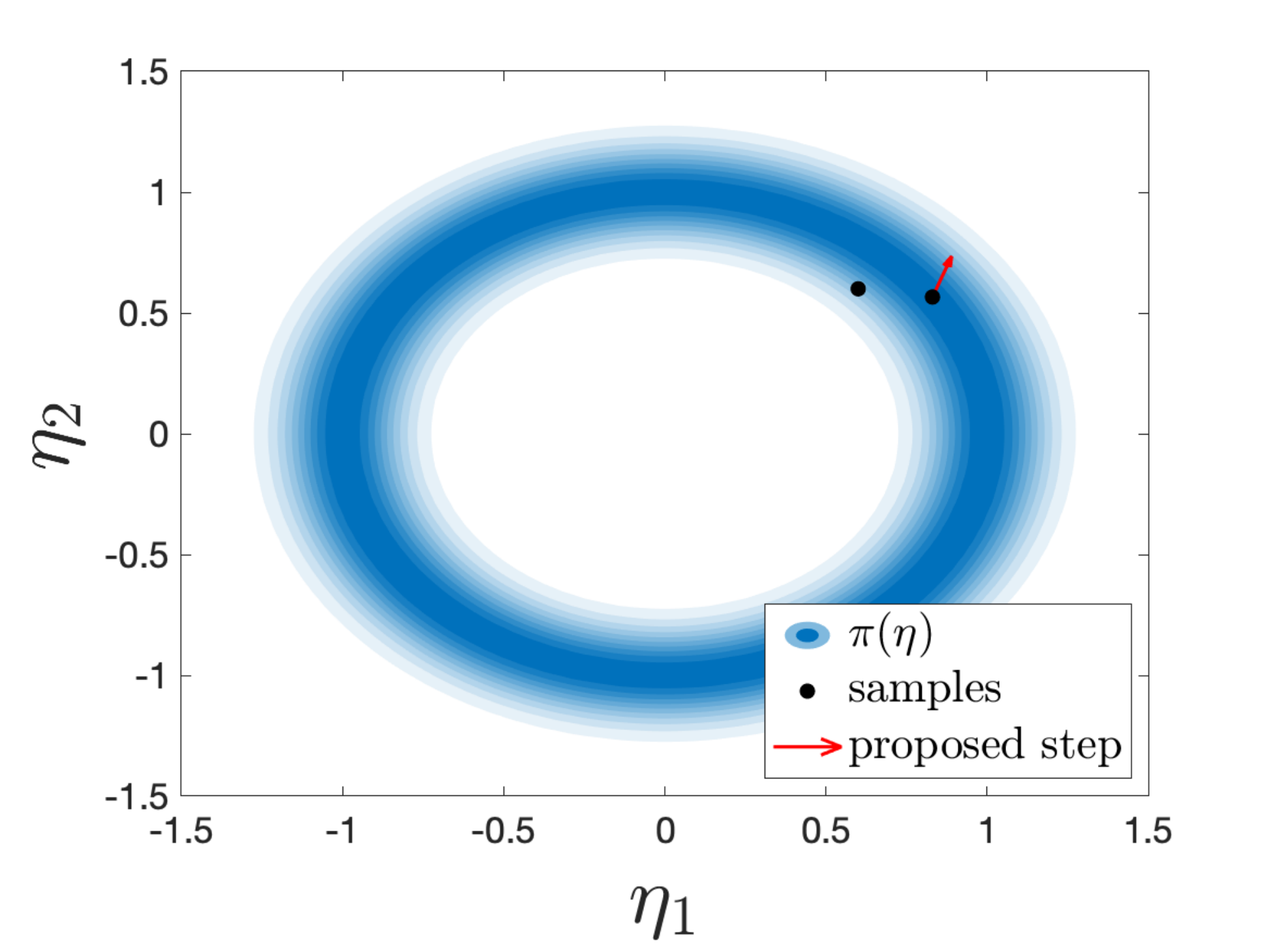}} \\
\vspace{5mm}
\subcaptionbox{MH $k=3$\label{fig:donut_MH_3}}
{\includegraphics[trim={10 5 45 25},clip,width=0.4\linewidth]{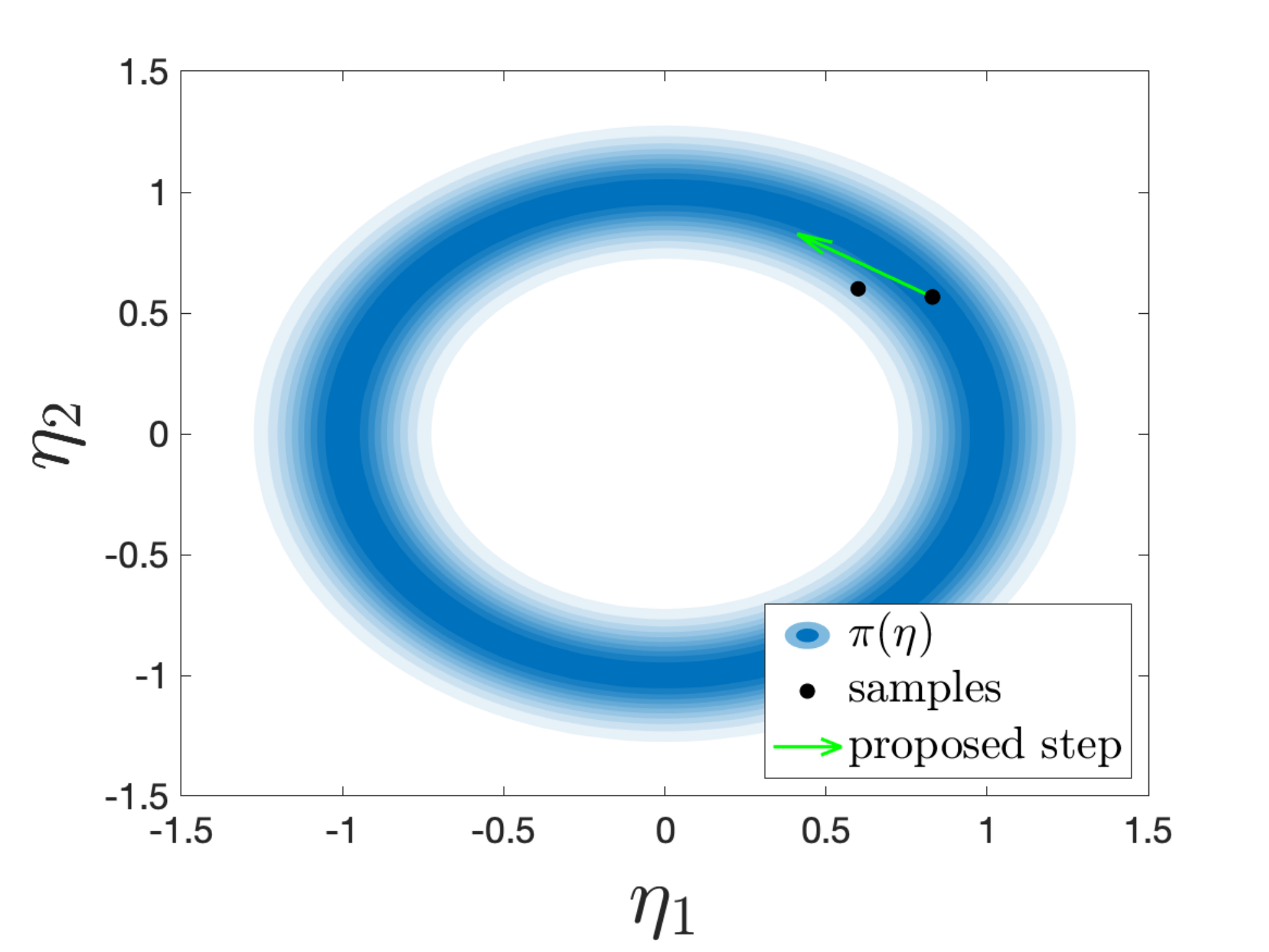}}
\subcaptionbox{MH $k=1$ to $10$\label{fig:donut_MH_1_10}}
{\includegraphics[trim={10 5 45 25},clip,width=0.4\linewidth]{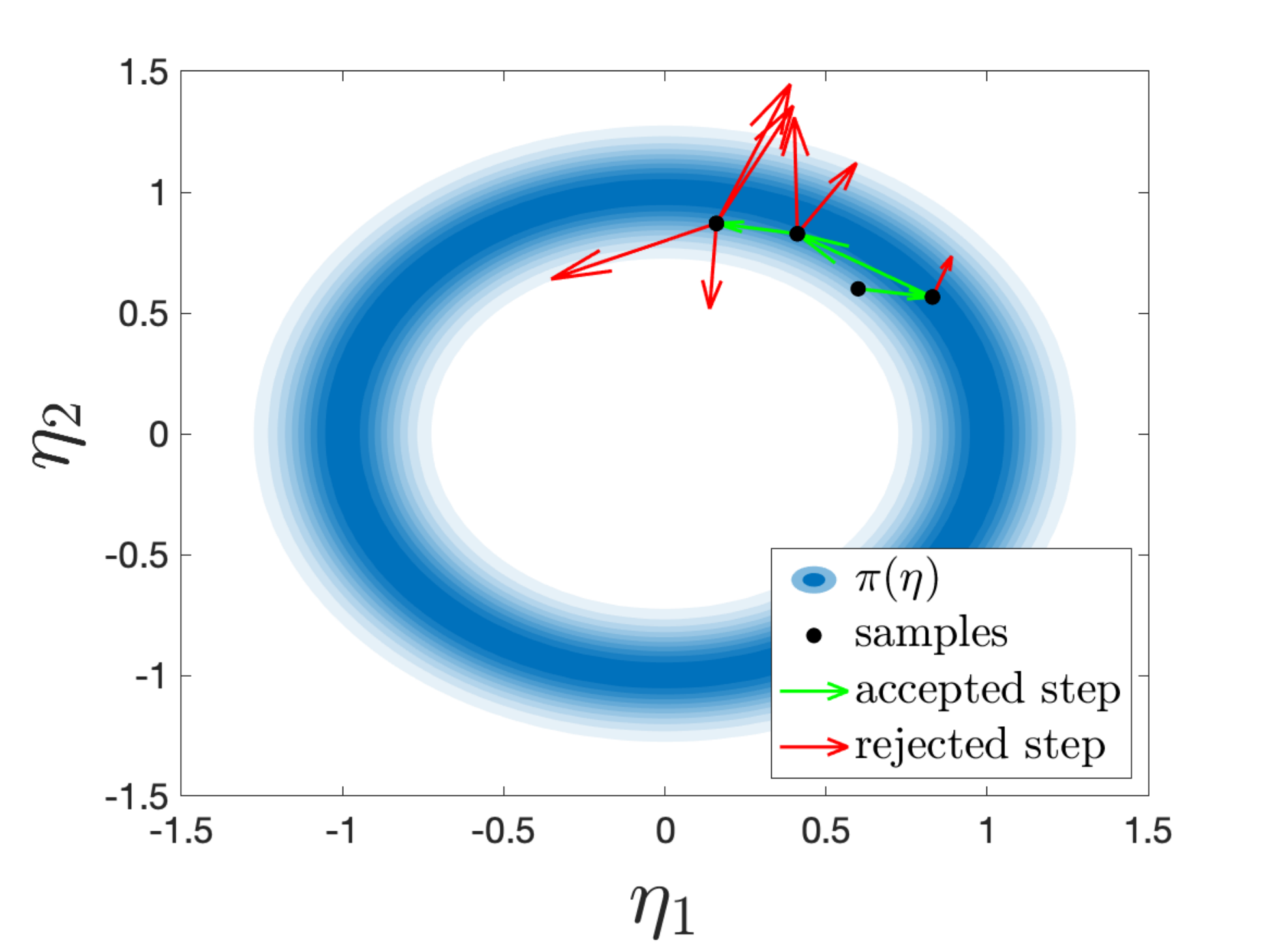}}

\caption{\em Illustration of MH to sample from a doughnut-shaped
  target in two-dimensions. Proposed steps are indicated by an arrow,
  with green indicating acceptance and red rejection }
\label{fig:donut_exampleMH}
\end{figure*}

The result for a random walk proposal with the MH algorithm are shown
in Figure~\ref{fig:donut_exampleMH}, where clearly little movement over the first 10
iterations has occurred and most moves have been rejected.

\begin{figure*}[htb]
\centering
\subcaptionbox{HMC $k=1$\label{fig:donut_HMC_1}}
{\includegraphics[trim={10 5 45 25},clip,width=0.4\linewidth]{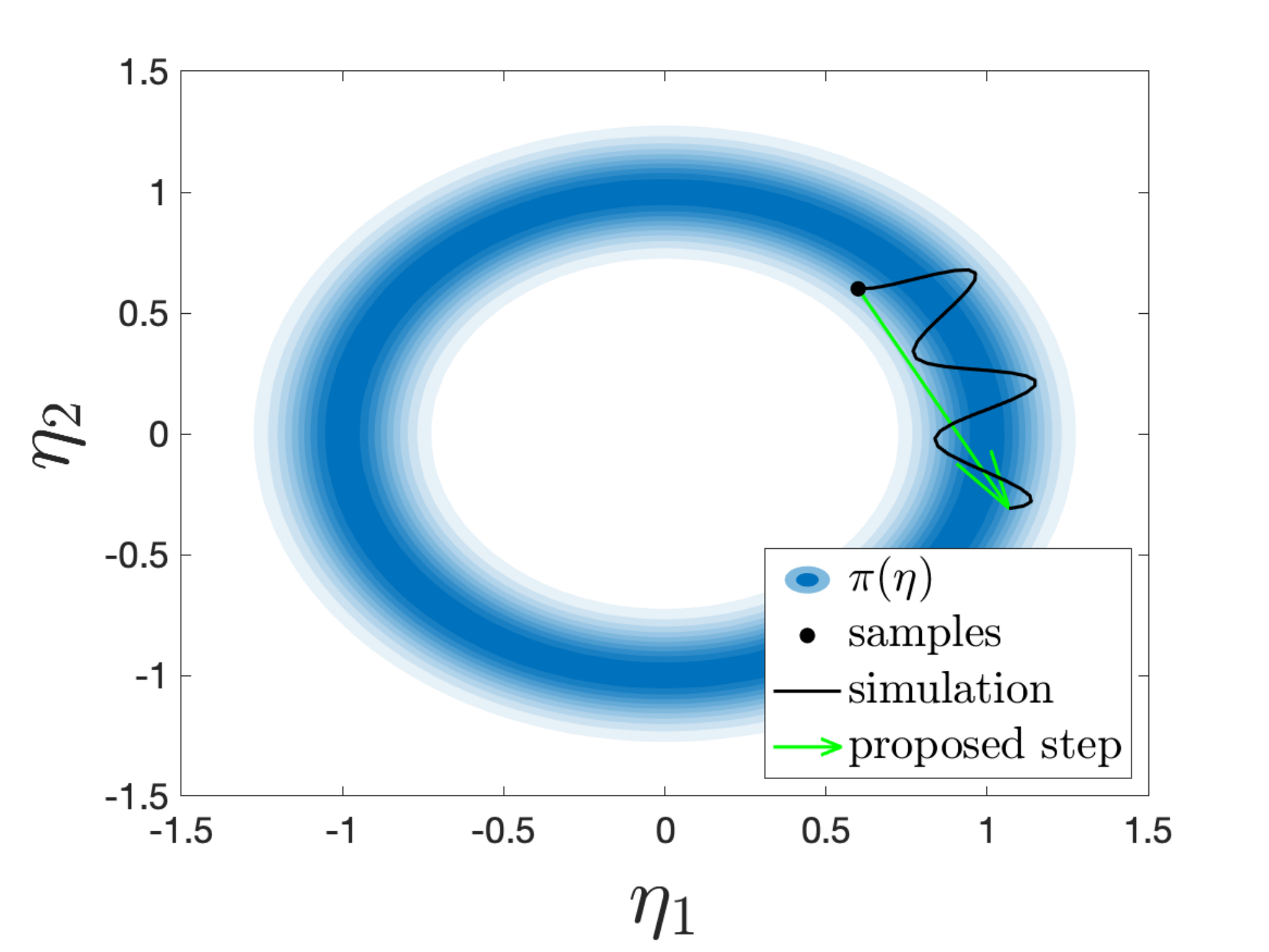}}
\subcaptionbox{HMC $k=2$\label{fig:donut_HMC_2}}
{\includegraphics[trim={10 5 45 25},clip,width=0.4\linewidth]{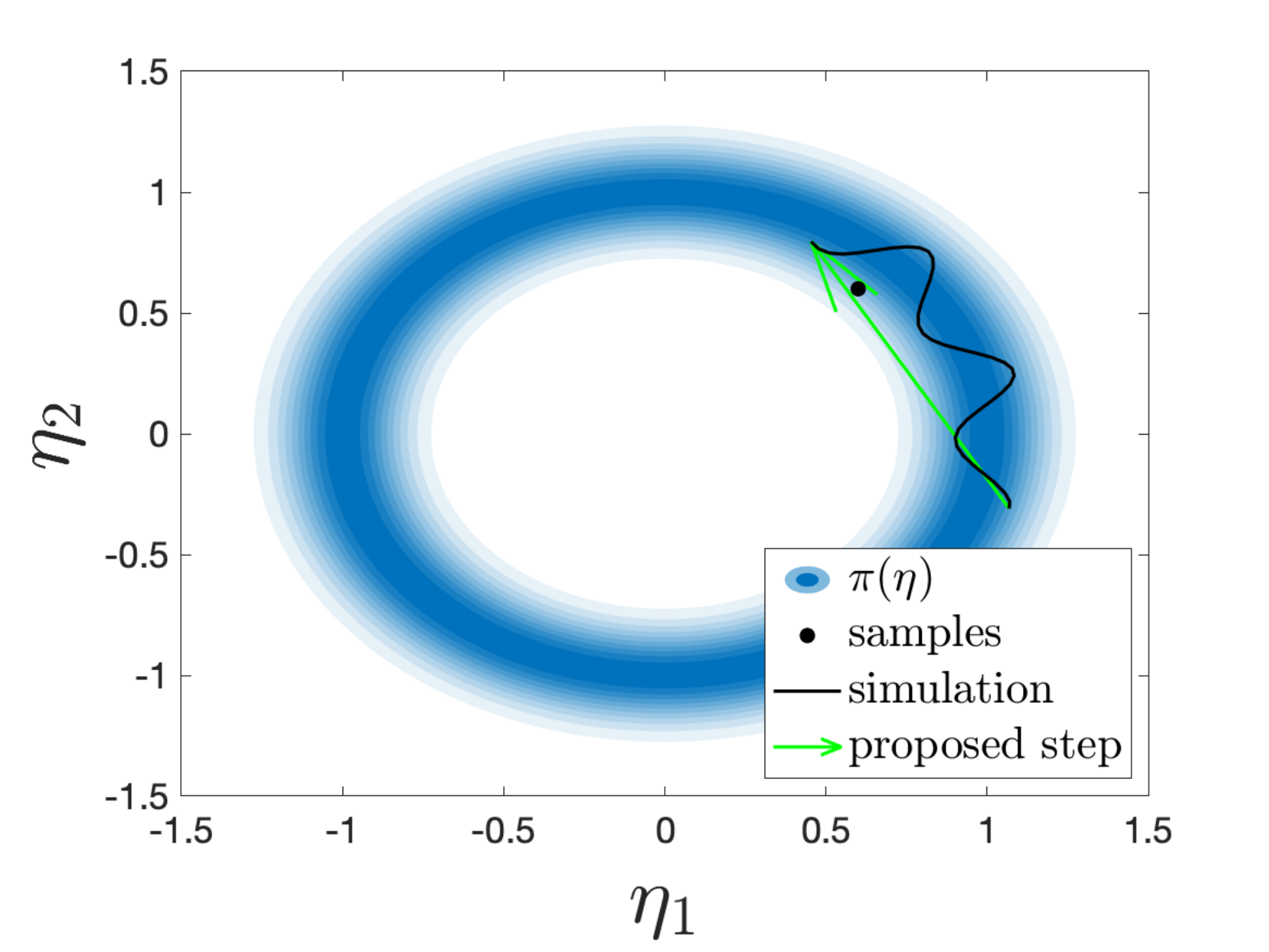}} \\
\vspace{5mm}
\subcaptionbox{HMC $k=3$\label{fig:donut_HMC_3}}
{\includegraphics[trim={10 5 45 25},clip,width=0.4\linewidth]{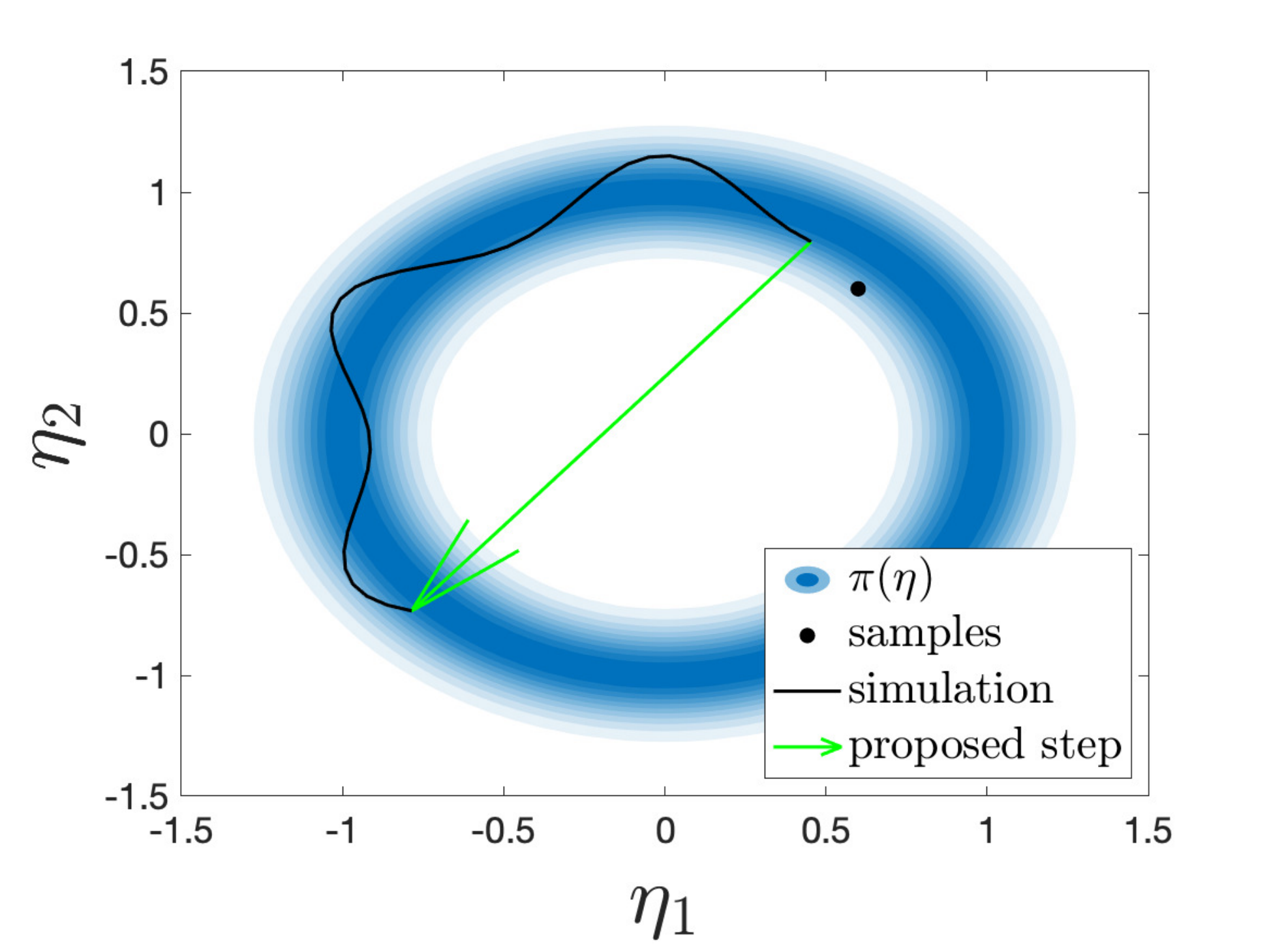}}
\subcaptionbox{HMC $k=1$ to $10$\label{fig:donut_HMC_1_10}}
{\includegraphics[trim={10 5 45 25},clip,width=0.4\linewidth]{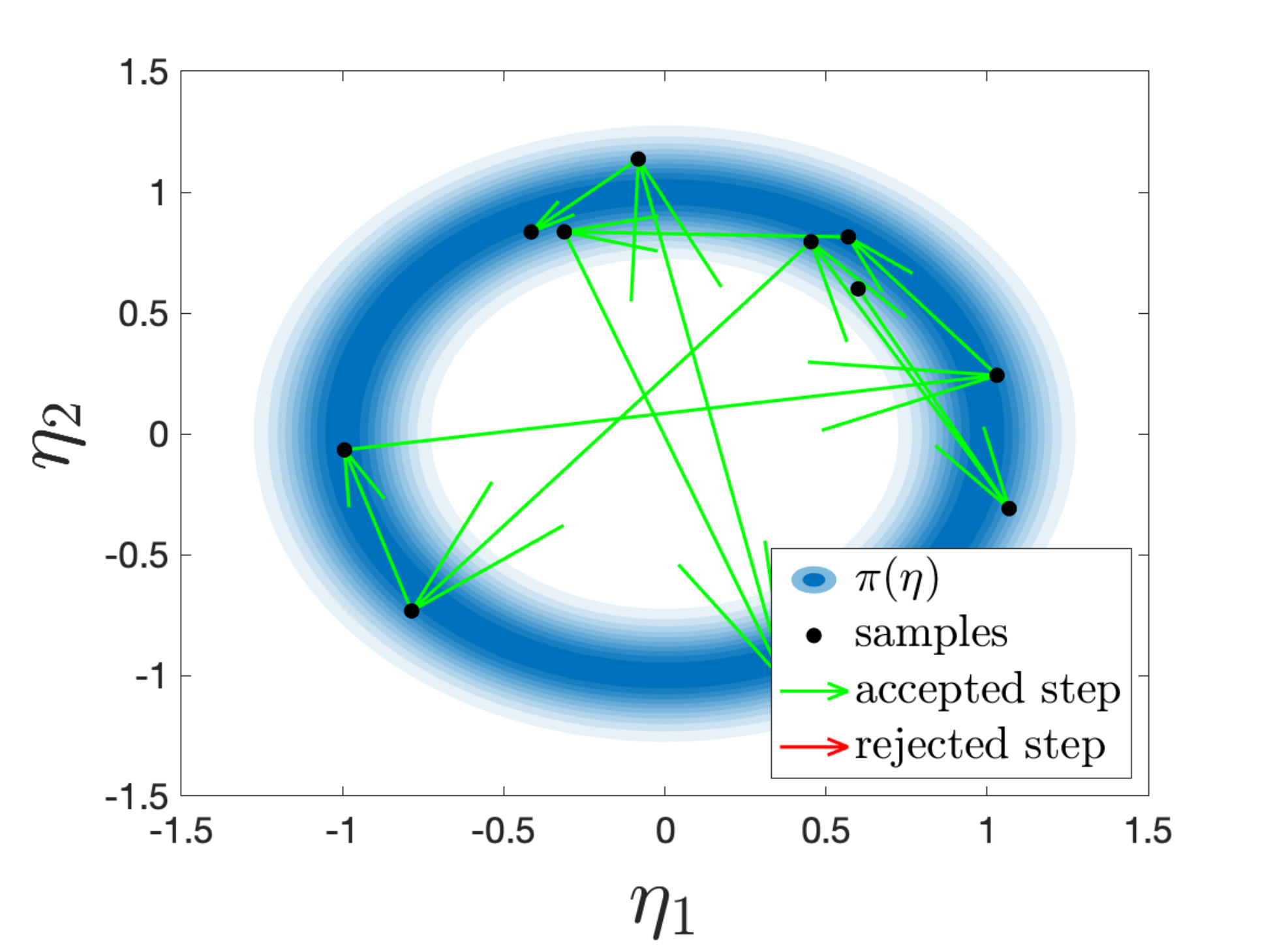}}
\caption{\em Illustration of HMC to sample from a doughnut-shaped
  target in two-dimensions. Proposed steps are indicated by an arrow,
  with green indicating acceptance and red rejection - there are no
  rejections.  Black lines illustrate the trajectory of $\eta(t),
  t\in[0,L]$ over the step of simulating the Hamiltonian dynamics.}
\label{fig:donut_exampleHMC}
\end{figure*}

By way of contrast, the moves for the HMC method over the first 10
iterations are shown in Figure~\ref{fig:donut_exampleHMC} and show all moves accepted with
much better exploration of this doughnut support target.

A short video showing the evolution of this HMC Markov chain is available at
\url{https://youtu.be/cyqh4mUB3bc}.

\section{Convergence}
\label{sec:convergence}

To this point, it has only been established that the HMC method
implements a Markov chain for which the joint target
$\widetilde{\pi}(\eta,\rho)$ is an invariant density of the chain with
marginal over $\rho$ satisfying \eqref{eq:12}.  Being an invariant
density it is then possible that the distribution of realisations
will converge to $\widetilde{\pi}$, but to this point there is no guarantee this
will actually happen, and indeed it has not been established that
$\widetilde\pi$ is unique as an invariant density.

Taking this latter question of uniqueness first, and to streamline
notation set
\begin{equation}
\label{eq:23}
\lambda \triangleq (\eta,\rho), \quad \dim \lambda = n_{\lambda}
\end{equation}
and define
\begin{equation}
\Lambda \triangleq \left\{\lambda\in\mathbf{R}^{n_{\lambda}} :
                     \widetilde{\pi}(\lambda) > 0\right\},
\label{one}
\end{equation}
\begin{equation}
\Gamma \triangleq \left\{\lambda\in\mathbf{R}^{n_{\lambda}} :
q(\lambda) > 0\right\} \label{eq:41}
\end{equation}
where $q(\lambda^{\prime}) = q(\eta^{\prime},\rho^{\prime})$ is the
proposal density achieved by the HMC approach of
Algorithm~\ref{alg:hmc} of randomly drawing
$\rho^{\prime}\in\mathcal{N}(0,\mathcal{M})$ and then simulating the
Hamiltonian dynamics to deliver a proposed realisation
$\eta^{\prime}$.

It is clear that the HMC method can only effectively sample the entire
support $\Lambda$ of the target $\widetilde\pi$ if
\begin{equation}
\label{eq:42}
\Lambda \subseteq \Gamma.
\end{equation}
In fact, if this holds $\widetilde\pi$ is the unique invariant
density.
\begin{lemma}
\label{m:1}
Suppose that $\Lambda$ is connected, that \eqref{eq:42} holds and that
$\widetilde\pi(\lambda)$ is bounded over $\lambda\in \Lambda$. Then
$\widetilde\pi(\lambda)$ is the unique invariant density of the Markov chain
realised by Algorithm~\ref{alg:hmc}.
\end{lemma}
\begin{pf}
The transition kernel $K(\xi\mid\lambda)$ for the Markov chain
realised by Algorithm~\ref{alg:hmc} is given by
\begin{eqnarray}
K(\lambda^{\prime}\mid \lambda) &=& \alpha(\lambda^{\prime}\mid\lambda)q(\lambda^{\prime})\nonumber\\
&=& \min\left\{
q(\lambda^{\prime}), \frac{\widetilde{\pi}(\lambda^{\prime})}{\widetilde{\pi}(\lambda)}
\,q(\lambda^{\prime})
\right\}.
\end{eqnarray}
Therefore, under the assumption~\eqref{eq:42} it holds that
\begin{equation}
\label{eq:44}
K(\lambda^{\prime}\mid\lambda) > 0 \quad \forall \lambda^{\prime},\lambda\in\Lambda.
\end{equation}
Define the distribution $\Pi(A)$ associated with the invariant density
$\widetilde\pi(\lambda)$ according to
\begin{equation}
\label{eq:45}
\Pi(A) = \int_A \widetilde\pi(\lambda)\,{\rm d}\lambda
\end{equation}
and define the conditional distribution $P(A\mid \lambda)$ similarly:
\begin{equation}
\label{eq:46}
P(A\mid \lambda) = \int_A K(\lambda^{\prime}\mid\lambda)\,{\rm d}\lambda^{\prime}.
\end{equation}
Then via ~\eqref{eq:44} 
\begin{equation}
\label{eq:47}
\Pi(A) > 0 \Rightarrow P(A\mid\lambda) > 0
\end{equation}
and hence the Markov chain produced by Algorithm~\ref{alg:hmc} is
$\Pi$-irreducible since the above implies the chain can visit any part of the domain of
$\widetilde\pi$ infinitely often.  Hence, by Proposition 10.1.1
of~\cite{MR95j:60103} the chain is recurrent, and hence by Theorem
10.2.1 of~\cite{MR95j:60103}, the invariant distribution $P(A\mid\lambda)$, and
hence (by boundedness) the invariant density $\widetilde\pi(\lambda)$ are unique.
\end{pf}
This establishes that if the chain converges in a distributional
sense, then the target density $\widetilde\pi$ and its associated
distribution
\begin{equation}
\label{eq:48}
\Pi(A) = \int_A \widetilde\pi(\lambda)\,{\rm d}\lambda
\end{equation}
is the only possibility.  That this convergence actually occurs can
also be established.
\begin{lemma}
\label{m:2}
Under the assumptions of Lemma~\ref{m:1} and defining
$P^n(A\mid\lambda)$ as the distribution of the realisation of
Algorithm~\ref{alg:hmc} at the $n$'th iteration, it holds that 
\begin{equation}
\label{eq:49}
\lim_{n\rightarrow\in}\sup_{A\in\mathbf{R}^{n_{\lambda}}}\left|
P^n(A\mid\lambda_0) - \Pi(A)
\right|
= 0
\end{equation}
for any $\lambda_0\in\mathbf{R}^{n_{\lambda}}$.
\end{lemma}
\begin{pf}
The Markov chain implemented by Algorithm~\ref{alg:hmc} was
established in the proof of Lemma~\ref{m:1} to be positive recurrent
with invariant distribution $\Pi(A)$.  Furthermore, via~\eqref{eq:47}
it is aperiodic for any starting $\lambda_0$, and by Corollary
2 of~\cite{Tierney1994} it is also Harris-recurrent.
The result then follows via
Proposition 13.0.1 in ~\cite{MR95j:60103}.
\qed
\end{pf}
Finally, we note that while this distributional convergence is a goal
in designing Algorithm~\ref{alg:hmc}, ultimately its utility is in
estimating expected values of function with respect to the target
density via sample averages as per~\eqref{eq:3},~\eqref{eq:4}.  This
required convergence is established in the following Theorem.
\begin{theorem}
\label{h:1}
Under the assumptions of Lemma~\ref{m:1} and with the further
assumption that the proposal $q(\lambda^{\prime})$ implemented in
Algorithm~\ref{alg:hmc} satisfies $q(\lambda^{\prime})<
\infty\,\forall \lambda^{\prime}\in\mathbf{R}^{n_{\lambda}}$ and that
$\varphi:\mathbf{R}^{n_{\eta}}\rightarrow\mathbf{R}$
satisfies
\begin{equation}
\label{eq:50}
\int_{\mathbf{R}^{n_{\eta}}} |\varphi(\eta)|\pi(\eta)\,{\rm d}\eta < \infty
\end{equation}
then
\begin{equation}
\label{eq:51}
\lim_{M\rightarrow\infty}
\frac{1}{M}
\sum_{k=1}^M \varphi(\eta_k) =
\int_{\mathbf{R}^{n_{\eta}}} \!\!\!\varphi(\eta)\pi(\eta)\, {\rm
  d}\eta
\end{equation}
with probability one.
\end{theorem}
\begin{pf}
It has already been established in the proof of Lemma~\ref{m:2} that
the Markov chain realised by Algorithm~\ref{alg:hmc} is Harris
recurrent.  The result then follows by Theorem 17.1.7 of~\cite{MR95j:60103}.
\qed
\end{pf}

Note how central the assumption~\eqref{eq:42} is to all these
convergence results.  One of the drawbacks of the HMC method is that
it is usually not straightforward to establish that ~\eqref{eq:42}
holds in practice since it will critically depend on the Hamiltonian
simulation duration $L$.

If it is too short, then~\eqref{eq:42} may not hold.  If too long then
an uneccessary computional burden ensues.  This has led some authors
to suggest some specifics around the simulation step, such as randomising
the simulation duration $L$~\cite{BouRabee}.

\section{HMC for Bayesian Dynamic System Identification} 
\label{sec:application_of_hmc_to_system_identification}

We now discuss the application of the HMC method to the dynamic system
identification problem formulated in
section~\ref{sec:problem-description} at the beginning of this paper.
This involves using HMC with target
\begin{equation}
\label{eq:29}
\pi(\eta) = p(\eta\mid y_{1:N})
\end{equation}
the posterior density of the vector $\eta$ given the observed data $y_{1:N}$.

Instrumental to achieving this is the use of Bayes' rule to decompose
the posterior as
\begin{equation}
\label{eq:10}
p(\eta\mid y_{1:N}) = \frac{p(y_{1:N}\mid\eta)p(\eta)}{p(y_{1:N})}
\end{equation}
where $p(y_{1:N}\mid\eta)$ is the likelihood of the observed data for
a given $\eta$, $p(\eta)$ is a prior for $\eta$, and $p(y_{1:N})$ is
an $\eta$-independent constant that ensures the area under the
posterior density $p(\eta\mid y_{1:N})$ is one.

In turn, by repeated applications of Bayes' rule the likelihood can be decomposed as
\begin{equation}
\label{eq:14}
p(y_{1:N}\mid\eta) = p(y_0)\prod_{t=1}^N p(y_t\mid y_{1:t-1},\eta).
\end{equation}
The computation of the ``one step ahead prediction density''
$p(y_t\mid y_{1:t-1},\eta)$ then depends on the model structure
chosen.

For example, for the general linear transfer function model structure
\begin{equation}
\label{eq:19}
y_t=G(q,\theta)u_t +H(q,\theta)e_t =\frac{A(q,\theta)}{B(q,\theta)}u_t+\frac{C(q,\theta)}{D(q,\theta)}e_t
\end{equation}
where $A(q,\theta),\cdots,D(q,\theta)$ are polynomials in the shift
operator $q$ and parametrised by the vector $\theta$, while $\{e_t\}$
is an i.i.d.\ sequence of zero mean random variables with variance
$\sigma_e^2$ it is possible to rewrite it as 
\begin{equation}
\label{eq:21}
y_t = GH^{-1}u_t + [1-H^{-1}]y_t+e_t
\end{equation}
If $C(q,\theta)$ and $D(q,\theta)$ are constrained to be monic
(zero'th order constant term in the polynomials fixed at $1$) then
$[1-H^{-1}]y_t$ is purely a function of $y_{1:t-1}$ with no dependence
on $y_t$ and hence with the definition of the one step ahead predictor
\begin{equation}
\label{eq:27}
\widehat{y}_{t\mid t-1} = GH^{-1}u_t + [1-H^{-1}]y_t
\end{equation}
then if $e_t\sim p_e(\cdot)$ the likelihood becomes
\begin{equation}
\label{eq:28}
p(y_{1:N}\mid\eta) = p(y_0)\prod_{t=1}^N p_e(y_t-\widehat{y}_{t\mid
  t-1}), \quad \eta^{T} = [\theta^T,\sigma_e].
\end{equation}
Similarly, in the case of linear, possibly time varying state space
model structure
\begin{equation}
\label{eq:ssm}
\left [ \begin{array}{c}x_{t+1}\\ y_t \end{array}\right ]
= \left [ \begin{array}{cc} A_t(\theta)&B_t(\theta)\\C_t(\theta)&D_t(\theta)\end{array}\right ]
\left [\begin{array}{c}x_t\\u_t\end{array}\right] + \left [
\begin{array}{c}\nu_t\\e_t\end{array} \right ].
\end{equation}
where $x_t\in\mathbf{R}_{n_x}$ is a state vector and $A_t(\theta)\cdots
D_t(\theta)$ are matrices with dimensions conformal to the above and
the dimensions of the possibly multivariable (in this case) $y_t$ and
$u_t$, then if the i.i.d.\ random variables $\nu_t$ and $e_t$
in~\eqref{eq:ssm} are Gaussian distributed
\begin{equation}
\label{eq:30}
\left[
\begin{array}{c}
\nu_t\\u_t
\end{array}
\right]\sim \mathcal{N}(0,\Sigma_t)
\end{equation}
then the likelihood is again given by~\eqref{eq:28} with
$\widehat{y}_{t\mid t-1}$ computed using the standard Kalman filter
recursions and $\eta^T=[\theta^T,\mbox{vec}\{\Sigma\}^T]$. 

In the linear but non-Gaussian case, or the non-linear state space
case the one-step ahead predictor density can still be formulated 
\begin{equation}
\label{eq:31}
p(y_t\mid y_{1:t-1},\eta) = \int p(y_t | x_t, \eta) p(x_t |
y_{1:t-1},\eta)\,{\rm  d}x_t
\end{equation}
but the associated multi-dimensional integral calculation is generally
intractable.

A strategy to deal with this problem is to include the full state
history $x_{1:N}$ in the parameter vector, so that if $\theta$
encompasses the remaining components in $\eta^T =
[\theta^T,\myVec{\{x_{1:N}\}}]$ then again using Bayes' rule
\begin{equation}
\label{eq:32}
\begin{split}
\pi(\eta) &= p(\theta,x_{1:N}\mid y_{1:N}) \\
&\propto
p(y_{1:N}\mid \theta,x_{1:N})p(x_{1:N}\mid \theta)p(\theta).
\end{split}
\end{equation}
Depending on the model structure, these two component may well be
tractable to compute.  For example, for the possible nonlinear state space
structure
\begin{equation}
\label{eq:33}
\begin{split}
x_{t+1} &= f(x_t,u_t) + w_t,\\
y_t&=g(x_t,u_t)+e_t,
\end{split}
\end{equation}
with 
\begin{equation}
\label{eq:36}
\quad w_t\sim p_w(\cdot),\quad\,e_t\sim p_e(\cdot)
\end{equation}
then it is straightforward to evaluate:
\begin{equation}
\label{eq:34}
\begin{split}
p(y_{1:N}\mid \theta,x_{1:N}) &= \prod_{t=1}^N p(y_t\mid x_t,\theta)\\
& =
\prod_{t=1}^N p_e(y_t-g(x_t,u_t))
\end{split}
\end{equation}
\begin{equation}
\label{eq:35}
\begin{split}
p(x_{1:N}\mid\theta) &= \prod_{t=1}^N p(x_t\mid x_{t-1},\theta) \\
&=\prod_{t=1}^N p_w(x_t-f(x_{t-1},u_{t-1})).
\end{split}
\end{equation}
After computing realisations from the joint posterior
$p(\theta,x_{1_T}\mid y_{1:t})$ we can marginalise over $x_{1:N}$
(i.e. ignore the realisations of $x_{1:N}$) to compute expectations
with respect to the posterior distribution of the model parameters
$p(\theta | y_{1:T})$.  

Note that vice-versa, if smoothed state estimates are also of
interest, then we can marginalise over the model parameters to compute
expectations with respect to the smoothing density $p(x_{1:N}\mid y_{1:N})$.

\subsection{Choice of Prior}
\label{sec:bayesian:prior}

The choice of prior $p(\eta)$ in~\eqref{eq:10} can be of
great benefit in achieving very effective dynamic system estimation
solutions.  Indeed, it is one of the great benefits of the Bayesian
approach that it provides a principled approach for injecting prior
knowledge, to then be combined with evidence from the data via the
likelihood $p(y_{1:N}\mid\eta)$ to deliver the final posterior
$p(\eta\mid y_{1:N})$.

For example, priors can be used to promote smoothness, sparsity,
stability and other properties of the estimated model and are closely
connected to the technique of ``regularisation'' in non-Bayesian
settings~\cite{PillonettoDinuzzoChenDeNicolaoLjung2014,PolsonScott2010}.

For example, it is very well known that Gaussian prior choice
\begin{equation}
\label{eq:37}
p(\eta) = \prod_{i=1}^{n_{\eta}}\mathcal{N}(0,\sigma_i^2)
\end{equation}
is closely related to imposing a 2-norm $\|\eta\|_2^2$ regularisation
penalty in an optimisation based (e.g. Maximum Likehood) estimation
approach.  Similarly with $\mathcal{L}$ denoting the Laplacian
density, the prior choice
\begin{equation}
\label{eq:37a}
p(\eta) = \prod_{i=1}^{n_{\eta}}\mathcal{L}(0,\sigma_i^2)
\end{equation}
implies a 1-norm $\|\eta\|_1^2$ regularisation penalty, and is
commonly used to  induce sparseness in the estimated $\eta$.

It is also possible, and indeed common to formulate a prior in a
hierarchical manner
\begin{equation}
\label{eq:38}
p(\eta) = p(\eta\mid \vartheta)p(\vartheta)
\end{equation}
where $\vartheta$ is referred to as a ``hyperparameter'', which has its
own prior attached to it as indicated above.

An example of this is the ``horseshoe
prior''~\cite{Carvalho,PolsonScott2010,PiironenVehtari2017} 
\begin{equation}
\label{eq:bayes:hs}
\begin{split}
	&p(\eta_i)\sim\mathcal{N}(0, \beta_i^2\tau^2),\\ 
	&\beta_i \sim \mathcal{C}^+(0, 1), \quad \tau \sim \mathcal{C}^{+}(0,1),
\end{split}
\end{equation}
where $\mathcal{C}^+(0, 1)$ denotes a zero-mean half-Cauchy (over the
positive reals) with scale $1$. Here, each $\beta_i$ is an independent
shrinkage parameter relating to each $\eta_i$ and $\tau$ is a global
shrinkage parameter.  This prior is again designed to promote
sparseness in the estimated vector $\eta$ in that elements will be
zero or close to it unless there is evidence from the data $y_{1:N}$
otherwise.  Although more complicated than the Laplace
prior~\eqref{eq:37a}  also designed to produce sparseness, it has been
observed to increase the accuracy of the resulting estimate~\cite{Carvalho}.

\subsection{Model Averaging}

The parameter posterior for hierarchical Bayesian models is still
computed using Bayes' rule.  However, we are now required to
marginalise the posterior over all the possible values of $\vartheta$
supported by the data, which corresponds to
\begin{align*}
	p(\eta | y)=\frac{p(y | \eta)}{p(y)}\dint p(\eta | \vartheta)p(\vartheta)\dx{\vartheta}.
\end{align*}
This is a type of model averaging where all possible choices of
$\vartheta$ are averaged over and therefore the influence of
$\vartheta$ is removed from the model.  As a result, Bayesian methods
are not prone to overfitting as no single value of $\vartheta$ is
selected.

Although exact marginalisation is intractable for many models of
interest, it can easily be performed in the case where we have samples
from the joint posterior
\begin{equation}
	\eta_k,\vartheta_k \sim p(\eta,\vartheta | y),
\end{equation}
by discarding the $\vartheta_k$ values. Sampling from the joint
posterior could be performed using the MCMC methods, such as MH or
HMC, discussed earlier

\section{Numerical Examples} 
\label{sec:numerical_examples}

In this section, we illustrate the use of HMC on a range of examples
from system identification. These simulations demonstrate that HMC
provides a practical means for sampling from the required posterior
distribution; this includes both classical model structures such as
ARX and output-error models, and more challenging nonlinear state-space
models.

Section~\ref{sub:autoregressive_exogenous} compares HMC with other
standard MCMC methods for estimation of an autoregressive exogenous
(ARX) model of known model
order. Section~\ref{sub:auto_regressive_exogenous_continued_} extends
the ARX example to estimation when the model order is unknown and
compares the effect of different priors when used within
HMC. Section~\ref{sub:output_error} demonstrates estimation of the
popular output error model from short data length and with unknown
model order using HMC and compares the results with regularised
maximum likelihood
estimates. Section~\ref{sub:identification_with_measurement_outliers}
demonstrates the estimation an ARX model when the measurements contain
significant outliers by utilising a Student's T distribution on the
errors and highlighting the flexibility of HMC. Finally,
section~\ref{sub:nonlinear_inverted_pendulum} showcases the joint
state and parameter estimation for a non-linear rotary inverted
pendulum from real data.

Depending on what is appropriate the estimates are assessed by; comparison to the true parameters, comparing the true and estimated frequency response, and/or using the model fit metric \cite{Ljung1999} given by
\begin{align}\label{eq:MF}
	\mathsf{MF}=100\left(1-\frac{\displaystyle \sum_{t=1}^T \left( \widehat{y}_t - y_t \right)^2}{\displaystyle \sum_{t=1}^T y_t^2}
	\right),
\end{align}
where $\widehat{y}_t$ and $y_t$ denote the one-step-head predictor and the observation at time $t$, respectively.

The implementations are made in MATLAB and Python using the PyStan \cite{stan2018pystan} interface to Stan \cite{carpenter2017stan}. The complete source code for these examples is provided on GitHub at \url{https://github.com/jnh277/hmc_lin_sys}.

\subsection{Autoregressive exogenous} 
\label{sub:autoregressive_exogenous}
This section provides an example of identifying an ARX model from data. It compares the ability of MH, mMALA, and HMC to provide uncorrelated samples from the posterior density $\eta \sim p(\eta | y_{1:T})$ using the model given by
\begin{align}\label{eq:arxmodel}
	y_t=\sum_{k=1}^{n_a} a_k y_{t-k}+\sum_{k=0}^{n_b} b_k u_{t-k}+e_t,
\end{align}
where $e_t \sim \mathcal{N}(0,\sigma_e^2)$ denotes zero-mean additive Gaussian noise. 
Here, $y_{1:T}$ and $u_{1:T}$ denotes the output and input, respectively.
Finally, the parameters of the model to identify using input-output data are $\eta=\{a_{1:n_a}, b_{0:n_b}, \sigma_e\}$.

The ARX model is simulated for $T=1000$ data points with order $n_a = n_b=2$ and parameters $\{-1.5, 0.7, 0, 1, 0.5, 1\}$ using a random binary input.
Moreover, we assume that the true model order is known to simplify the inference.
For inference, $667$ data points are used; with the remainder reserved for validation.
The \texttt{arx} command in MATLAB is used to obtain the maximum likelihood estimate.

\begin{figure}[tb]
	\centering
	\includegraphics[width=1.0\linewidth]{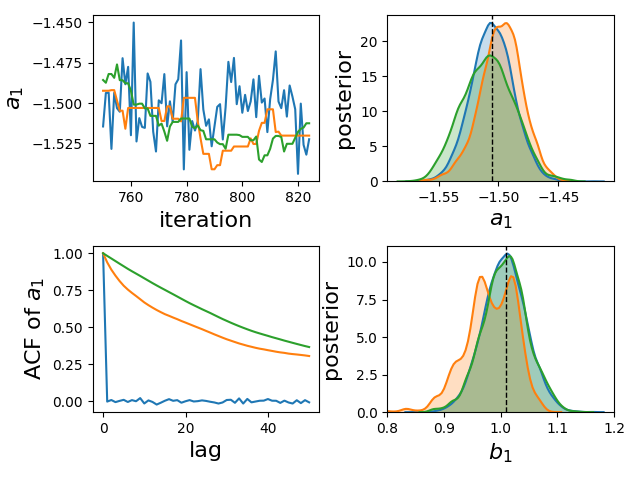}
	\caption{System identification of an ARX model \eqref{eq:arxmodel} of known order. The trace plots (top left) and automatic correlation factor (bottom left) for $a_1$ as well as the posterior distributions of $a_1$ and $b_1$ obtained using MH (green), mMALA (orange), and HMC (blue). Also shown is the maximum likelihood estimate (dashed vertical line).}
	\label{fig:arx_example1_diagnostics}
\end{figure}

A Bayesian estimate is obtained using HMC, MH, and mMALA with an L2 (Gaussian) prior on the parameters. 
Figure~\ref{fig:arx_example1_diagnostics} shows the trace plots and automatic correlation factor for $a_1$ as well as the posterior marginal distributions of $a_1$ and $b_1$. 
The trace is a plot of the state of the Markov chain at each iteration. Note that the chain from MH and mMALA exhibit a much larger auto-correlation than the chain generated by HMC. 
This is also apparent as the ACF falls off quicker for HMC compared with the other two approaches. 
Finally, calculating the model fit for each estimate gives $96.09$ for Matlab's \texttt{arx} command, $96.08$ using HMC, $96.04$ using mMALA, and $94.03$ using MH.

Recall the discussion in the previous section, a large persistent
autocorrelation results in poor mixing and poor performance in the
MCMC method.  Even on this relatively low order example, the
difference in the autocorrelation factor is a strong indicator of the
greater efficiency of HMC sampling from the target distribution. This
difference becomes even more noticeable as the dimension of the
parameter space increases. Even for high dimensional parameter spaces,
HMC using the No-U-Turn sampling algorithm is typically able to
efficiently provide samples with minimal user tuning.  Having made
this comparison, the remainder of the examples in this section focus
on HMC and details relevant to its use for the model type considered.


\subsection{Auto-regressive exogenous, continued.} 
\label{sub:auto_regressive_exogenous_continued_}
After the previous sanity check, we proceed with a slightly more interesting example where the model orders are unknown.
In this setting, the standard approach to maximum likelihood inference would be to include a regularisation term or to use cross-validation to determine a suitable model order.
In the Bayesian setting, the standard approach is to use a sparseness prior which regularises the model and \emph{turns off} unnecessary parameters in an over-parameterised model. 
Using HMC, this example compares the effect of L1 (exponential) prior, L2 (Gaussian) prior, and the horseshoe prior.

\begin{figure}[tb]
	\centering
	\subcaptionbox{\label{fig:impact_of_prior_a}}{
	\includegraphics[width=1.0\linewidth]{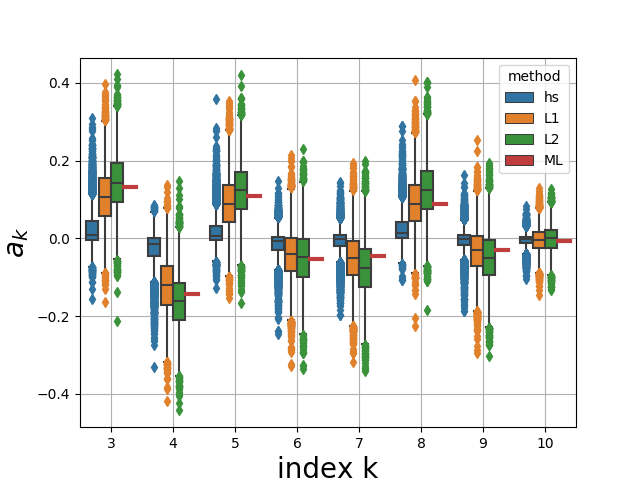}}
	\subcaptionbox{\label{fig:impact_of_prior_b}}{\includegraphics[width=1.0\linewidth]{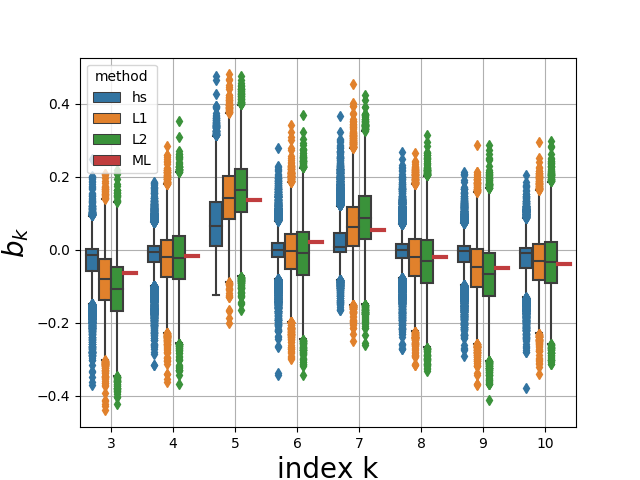}}
	\caption{System identification of an ARX model \eqref{eq:arxmodel}  of unknown order. Shown is the posterior distributions for the parameters that are not part of the true model, i.e. $a_k$ and $b_k$ for $k>2$. The Bayesian estimate using L1, L2, and horseshoe priors is shown, along with the maximum likelihood estimate using a tuned-correlated kernel for regularisation.}
	\label{fig:prior_choice}
\end{figure}

In this example, we use $n_a=n_b=10$ as a guessed maximum model order and compute the maximum likelihood estimate using \texttt{Matlab}'s \texttt{arx} function with the popular tuned-correlated (TC) kernel employed to regularise the results. The different impacts that each prior has on the posterior distribution of the parameters are evident in Figure~\ref{fig:prior_choice}. Shown are the marginal distributions for the parameters that are not part of the true model, i.e. for $k>2$, which indicates that the horseshoe prior is more effective at reducing the value of these parameters than the L1 or L2 priors. Also shown are the maximum likelihood estimates.

\begin{figure}[tb]
  \centering
  \includegraphics[width=0.9\linewidth]{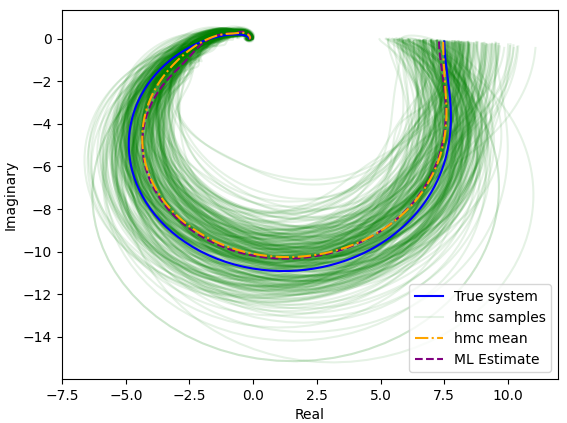}
  \caption{Estimated Nyquist diagram for the ARX model with unknown model order. The HMC estimate is shown for each sample from the posterior, indicating the uncertainty, as well as the conditional mean. Also shown is the true Nyquist diagram and the maximum likelihood estimate using a TC kernel for regularisation.}
  \label{fig:arx_unknown_order_nyquist}
\end{figure}

As well as promoting sparseness, it would be good if choosing the horseshoe prior still yielded accurate estimates of the parameters. 
The model fit given using ML is 89.3, and for HMC using L1, L2, and horseshoe priors are respectively 89.2, 89.1, and 89.4.
Additionally, the Nyquist diagram of the HMC samples with a horseshoe prior is shown in Figure~\ref{fig:arx_unknown_order_nyquist} and compared to the true systems and the ML estimate.

\subsection{Output error} 
\label{sub:output_error}
This section demonstrates the Bayesian estimation of an output error (OE) model using HMC. Consider the model 
\begin{align}
	y_t=\frac{b_0 + b_1 q^{-1} + \ldots + b_{n_b} q^{-n_b}}{1 + f_1 q^{-1} + \ldots + f_{n_f} q^{-n_f}}\,u_t+e_t,
	\label{eq:oemodel}
\end{align}
where the measurements are corrupted by additive Gaussian noise $e_t\sim\mathcal{N}(0,\sigma_e)$.
The parameters of the model to identify using input-output data $\{y_{1:T}, u_{1:T}\}$ are $\eta=\{b_{0:n_b}, f_{1:n_f}, \sigma_e\}$.

We simulate $T=200$ data points from the OE model with order $n_b=n_f=4$ and parameters $\{0.0, 0.024, 0.170, 0.113, 0.007, 1.22, 0.56, -0.18, 0.5\}$ using a square wave input. Half the data points are used for inference and the remainder are reserved for validation.

\begin{figure}[tb]
  \centering
  \includegraphics[width=0.9\linewidth]{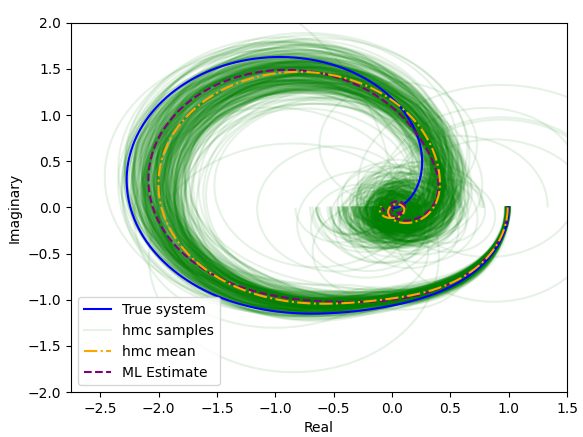}
  \caption{Nyquist diagram for the estimated output error models. 
  The HMC estimate is shown for each sample from the posterior and the conditional mean response.
  For comparison the maximum likelihood estimate and the true system are also shown.}
  \label{fig:oe_nyquist}
\end{figure}

With a guessed maximum model order of $n_b=n_f=10$, a Bayesian estimate of the model is given using HMC with a horseshoe prior, achieving a model fit of 99.42, and a maximum likelihood estimate using Matlab's \texttt{arx} command with regularisation achieving a model fit of 99.45. The corresponding Nyquist diagrams are shown in Figure~\ref{fig:oe_nyquist}. 

\subsection{Identification with measurement outliers} 
\label{sub:identification_with_measurement_outliers}
So far, the examples shown have only considered additive Gaussian noise.
This assumption, however, is problematic if the measurements include outliers.

The assumption of additive Gaussian noise is not required for Bayesian inference using HMC, and a large variety of noise models can be easily implemented.
This section revisits the example given in Section~\ref{sub:auto_regressive_exogenous_continued_} and considers noise that has been simulated from a Student's T distribution, $e_t\sim\mathcal{T}\left(\nu,0,\sigma_e^2\right)$. 
For low values of $\nu$ (the degrees of freedom), the Student's T distribution has fatter tails than a Gaussian distribution and so the measurements can have significant outliers.

In this example, we use $n_a = n_b = 10$ as a guessed model order and compare  the results from Matlab's \texttt{arx} command with TC regularisation against estimating the model using HMC and a Student's T distribution for the noise. Both the standard deviation and the degrees of freedom of the noise are estimated. Since for $\nu > 10$, the Student's T distribution becomes practically Gaussian, hence using this model is not restrictive. For the parameters, a horseshoe prior is used. This gives the parameters to be estimated using HMC as $\eta = \{a_{1:n_a},b_{0:n_b},\sigma_e,\nu\}$, where $\nu$ is constrained within $[1,\infty]$ and is given a Gamma prior, $\mathcal{G}(2,0.1)$.

\begin{figure}[tb]
  \centering
  \includegraphics[width=0.9\linewidth]{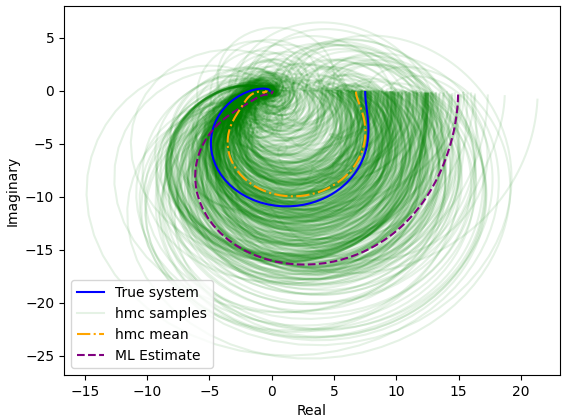}
  \caption{Nyquist diagrams of the estimated ARX models from data with measurement outliers. The Nyquist diagram for the true system (black), ML estimate (dashed purple), samples from the posterior $\eta\sim p(\eta | y_{1:T})$ (green), and the posterior mean (dashed orange) are shown.}
  \label{fig:robust_noise_bode}
\end{figure}

The Nyquist diagrams corresponding to the estimated models are shown in Figure~\ref{fig:robust_noise_bode}. Despite, Matlab's \texttt{arx} method using a Huber style cost to improve performance when outliers are present it's performance is degraded for this data. In contrast, HMC provides good estimates despite the measurement outliers.

\subsection{Nonlinear inverted pendulum} 
\label{sub:nonlinear_inverted_pendulum}
Up to this point, the examples have all focussed on linear systems using simulated data. This section provides an example of estimating a non-linear state space system from experimental data. The platform used is the rotary inverted pendulum QUBE-Servo 2 of QUANSER\textsuperscript{\textcopyright}. 
The dynamics of this system can be modelled according to

\small
\begin{equation}\label{eq:pend_process}
\begin{split}
    &M(\alpha)\pmat{\ddot{\vartheta} \\ \ddot{\alpha}} + \nu(\dot{\vartheta},\dot{\alpha})\pmat{\dot{\vartheta} \\ \dot{\alpha}} = \pmat{\frac{k_m(V_m - k_m\dot{\vartheta})}{R_m} - D_r\dot{\vartheta} \\ -\frac{1}{2}m_pL_pg\sin(\alpha)-D_p\dot{\alpha}},\\
         &M(\alpha) \hspace{-0.75mm}=\hspace{-0.75mm} \pmat{m_pL_r^2\hspace{-0.5mm}+\hspace{-0.5mm}\frac{1}{4}m_pL_p^2(1\hspace{-0.5mm}-\hspace{-0.5mm}\cos(\alpha)^2)\hspace{-0.5mm}+\hspace{-0.5mm}J_r & \hspace{-2mm} \frac{1}{2}m_pL_pL_r\cos(\alpha) \\ \frac{1}{2}m_pL_pL_r\cos(\alpha) & J_p + \frac{1}{4}m_pL_p^2}, \\
    &\nu(\dot{\vartheta},\dot{\alpha}) \hspace{-0.75mm}=\hspace{-0.75mm} \pmat{\frac{1}{2}m_pL_p^2\sin(\alpha)\cos(\alpha)\dot{\alpha} & \hspace{-2mm} - \frac{1}{2}m_pL_pL_r\sin(\alpha)\dot{\alpha} \\
    - \frac{1}{4}m_pL_p^2\cos(\alpha)\sin(\alpha)\dot{\vartheta} & 0}, \\
\end{split}
\end{equation}
\normalsize
where $m_p$ is the pendulum mass, $L_r$, $L_p$ and the rod and pendulum lengths, $J_r$, $J_p$ are the rod and pendulum inertias, $R_m$ and $k_m$ are the motor resistance and constant, $D_p$ and $D_r$ are the pendulum and arm damping constants, respectively. 

Measurements of the two angles, $\theta$ and $\alpha$, and the motor
current $I_m$ made at $8\text{ms}$ intervals are available. These
measurements are modelled according to 
\begin{equation}\label{eq:pend_meas}
	y_t =  \pmat{\theta(t) & \alpha(t) & \frac{V_m(t) - k_m\dot{\theta}}{R_m}}^{\Transp}
\end{equation}
Given that the parameters $m_p$, $L_p$ and $L_r$ can easily be
measured accurately, we focus on estimating the parameters $J_r$,
$J_p$, $K_m$, $R_m$, $D_p$, and $D_r$ using the input-output data
$\{V_{m,t},y_t\}_{t=1}^T$.  To do this we choose as latent variables
$x(t) = \pmat{\theta(t) & \alpha(t) & \dot{\theta}(t)&
  \dot{\alpha}(t)}^{\Transp}$.
This allows us to write a state space model of the form
\eqref{eq:ssm}. Additionally, covariance between the process noise,
$e_t$, and measurement noise $w_t$ was considered giving the joint
likelihood as
\begin{equation}
	\pmat{x_{t+1}\\y_{t}} \sim \mathcal{N}\left(\pmat{f(x_{t},V_{mt}) \\ h(x_t,V_{mt})} ,\Pi = \pmat{Q & S \\ S^T & R} \right),
\end{equation}
where $f$ and $h$ are given by equations \eqref{eq:pend_process} and
\eqref{eq:pend_meas}, respectively, combined with an ODE solver. In
this case, the Runge-Kutta $4^\text{th}$ order method was used.

\begin{figure}[tb]
	\centering
	\includegraphics[width=1.0\linewidth]{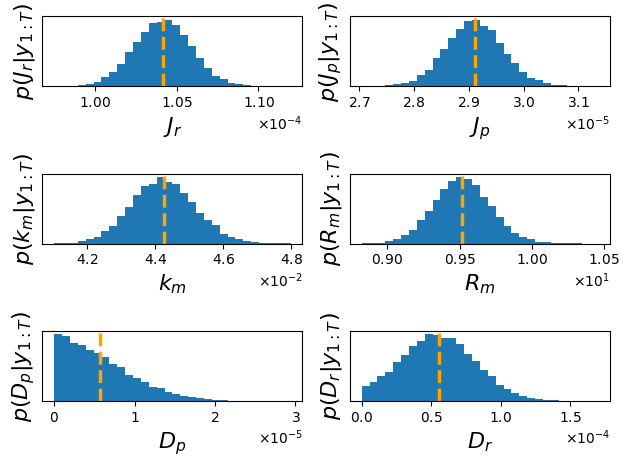}
	\caption{The marginal posterior distributions for the nonlinear inverted pendulum model parameters. The conditional mean value is indicated by a dashed orange line.}
	\label{fig:pendulum_params}
\end{figure}

HMC was used to infer the joint posterior of $x_{1:T}$, $J_r$, $J_p$,
$K_m$, $R_m$, $D_p$, $D_r$, and $\Pi$ from an experimentally collected
data set containing $375$ input-output pairs.  The model parameters
were constrained to be positive with a horseshoe prior and an LKJ
prior on the covariance matrix.  The model parameters were initialised
with a value of 1 and the latent variables were initialised at the
measurement values, where appropriate. The marginal posterior
distributions for the parameters are shown in
Figure~\ref{fig:pendulum_params}.  From this we can see that the most
likely value for $D_p$ is close to zero as joints use magnetic
bearings and whereas $D_r$ has a non-zero value introduced by cabling
slightly interfering with the smooth movement of the arm rotation.

\begin{figure}[tb]
	\centering
	\includegraphics[width=1.0\linewidth]{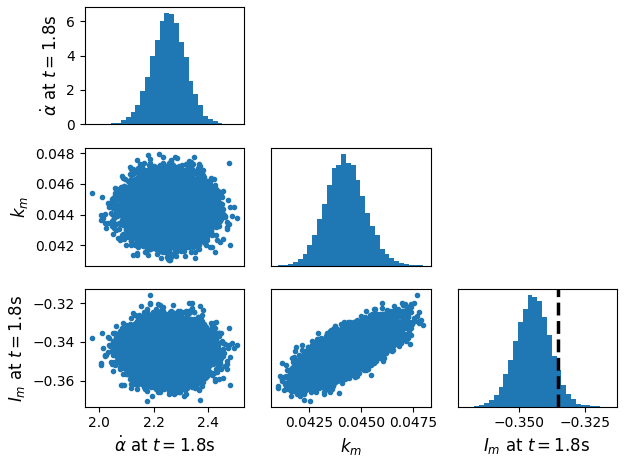}
	\caption{Posterior distributions of the parameters of the non-linear inverted pendulum system: $k_m$, $\dot{\alpha}$ at $t=1.8$s, and the predicted measurement of $I_m$ at $t=1.8$s. The joint distribution between each of pair is represented by the scatter plots. Also shown is the measured value for $I_m$ indicated by a dashed line.}
	\label{fig:pendulum_pairs}
\end{figure}

The estimates of the arm angles and angular velocities are compared to the measured values (and gradients of the measured values) in Figure~\ref{fig:pendulum_mean_states}. 
As HMC provides the full joint posterior of the states and the parameters there is far more information we can look at. 
Figure~\ref{fig:pendulum_pairs} gives examples of marginal and joint distributions between parameter-state, parameter-measurement, and state-measurement pairs. From this, for instance, we can see that there is correlation between the value for $k_m$ and the predicted measurements of $I_m$.

\begin{figure}[tb]
	\centering
	\includegraphics[width=1.0\linewidth]{images/pendulum_states.png}
	\caption{Smoothed state estimates for the nonlinear inverted pendulum. Shown is the mean estimate of the angles and angular velocities (dashed blue) compared to the measured values and gradient of measured values (black).}
	\label{fig:pendulum_mean_states}
\end{figure}

\section{Conclusion} 
\label{sec:conclusion}
This paper has presented Bayesian inference using Hamiltonian Monte Carlo within the context of system identification. 
Hamiltonian Monte Carlo is presented as a specific choice of the proposal for Metropolis-Hastings and its advantages over other proposals, such as mMALA, have been identified.
In particular, it allows for more efficient sampling from high dimensional parameter spaces with minimal user tuning.
This allows the application of HMC for Bayesian system identification to a broad range of model types.
These model types include classical models such as ARX and OE, through to nonlinear state-space models.
The ability to include the latent variables of a state-space as parameters to be sampled allows joint parameter estimation and state smoothing to be performed using HMC.






\bibliographystyle{plain}
\bibliography{references}

\end{document}